\documentclass[%
 reprint,
 nofootinbib,
 nobibnotes,
 bibnotes,
 amsmath,amssymb,
prb,
]{revtex4-2}

\usepackage{color,ulem}%
\usepackage{graphicx}
\usepackage{dcolumn}
\usepackage{bm}
\usepackage{hyperref}
\usepackage[mathlines]{lineno}

\begin{document}


\title{Linear and nonlinear Edelstein effects in Rashba superconductors}

\author{Hikaru Ueki}%
\author{Youichi Yanase}%
\affiliation{%
Department of Physics, Graduate School of Science, 
Kyoto University, Kyoto 606-8502, Japan
}%

\date{\today}

\begin{abstract}
We formulate a quasiclassical theory of the Edelstein effect in superconductors that incorporates both intraband and interband contributions. To describe the interband contribution, which is absent from the conventional leading-order quasiclassical formulation, we derive augmented Eilenberger equations in the presence of antisymmetric spin-orbit coupling. The intraband contribution is evaluated using multiband Eilenberger equations. We apply these formulations to supercurrent-induced surface spin magnetization in $s$-wave Rashba superconductors and investigate its dependence on temperature, distance from the surface, spin-orbit coupling strength, and supercurrent. The intraband contribution originates from a supercurrent-induced asymmetry of quasiparticles with opposite momenta and spin polarizations, whereas the interband contribution arises from the anomalous-velocity term generated by the momentum derivative of the Rashba spin-orbit potential. In the helicity basis, this anomalous-velocity term is expressed in terms of the Berry connection associated with the momentum dependence of the Rashba eigenstates. The intraband contribution increases linearly with the spin-orbit coupling strength, whereas the interband contribution exhibits a nonmonotonic dependence and is maximized when the Rashba spin splitting is comparable to the superconducting gap. Moreover, within the clean $s$-wave Rashba model considered here, we find that the nonlinear dependence of magnetization on the supercurrent arises solely from the interband contribution. Thus, although the intraband contribution dominates the linear Edelstein effect, the nonlinear Edelstein effect can serve as a useful probe of the interband contribution originating from quantum geometry.
\end{abstract}

\maketitle


\section{Introduction}
The Edelstein effect refers to the generation of spin magnetization by an electric current \cite{ede90}. It originates from spin-momentum locking induced by broken inversion symmetry and spin-orbit coupling. Its superconducting counterpart is of interest in the context of superconducting spintronics \cite{esc11,lin15}, because spin magnetization can be generated by dissipationless supercurrents rather than by dissipative currents in normal metals. Previous studies investigated the Edelstein effect in noncentrosymmetric $s$-wave superconductors \cite{ede95,yip02,yip05,oka06,kon15,he19,he20,dai24}, as well as in superconductors with mixed spin-singlet and spin-triplet pairing \cite{ede03,fuj05,hay06}. More recently, in $d$-wave Rashba superconductors, the magnetization was shown to be enhanced near surfaces hosting zero-energy states \cite{ike20}. However, the nonlinear dependence of magnetization on the supercurrent remains largely unexplored in superconductors. Such nonlinear responses, referred to as the nonlinear Edelstein effects, are important for understanding the magnetoelectric properties of noncentrosymmetric superconductors under strong supercurrents, particularly beyond the regime of validity of the linear response theory.

The standard Eilenberger equations \cite{eil68} provide a powerful framework for studying a wide range of superconducting phenomena, including nonlinear and disorder effects in the Meissner response \cite{yip92,xu95,sau22}, and they are applicable to spatially nonuniform states such as surface states \cite{buc95,buc95-2,fog97} and vortex states \cite{kle87,sch95,ich96,ich97}. The standard Eilenberger equations with Rashba spin-orbit coupling have also been formulated, and vortex-induced spin polarization, surface bound states, and equilibrium spin currents in noncentrosymmetric superconductors have been studied \cite{hay06,vor08,esc12}. However, these formulations retain only the leading-order terms in the quasiclassical expansion and therefore cannot describe effects arising from higher-order corrections. Examples include the equilibrium Hall effect and associated charge redistribution \cite{kit09} and, as demonstrated below, an interband contribution to the Edelstein effect in purely spin-singlet Rashba superconductors. Hayashi {\it et al.} calculated a magnetization perpendicular to both the circulating supercurrent and the magnetic field around a vortex in an $s+p$-wave Rashba superconductor using the standard Eilenberger equations \cite{hay06}. In their formulation, this magnetization arises from the spin-triplet $p$-wave component of the pair potential and vanishes in the purely spin-singlet $s$-wave limit. It is therefore distinct from the interband magnetization studied in this paper.

Charging effects and equilibrium Hall effects in superconductors can be described by augmenting the Eilenberger equations with the Lorentz-force term, the pair-potential-gradient term, and the term arising from the slope of the density of states \cite{kit09,uek18,mas19,jos20,ohu22}. These terms appear as next-to-leading-order contributions in the expansion of the Gor’kov equations \cite{gor59,gor60} with respect to the quasiclassical parameter $\delta\equiv 1/k_{\rm F}\xi_0$, where $k_{\rm F}$ is the Fermi wavenumber and $\xi_0$ is the coherence length. By contrast, the standard Eilenberger equations with Rashba spin-orbit coupling retain only the leading-order terms in this expansion \cite{hay06,vor08,esc12}. Consequently, they neglect terms involving the momentum derivative of the Rashba spin-orbit potential, which appear at next-to-leading order in the quasiclassical expansion and have also been incorporated into normal-state transport equations \cite{she18}. This momentum derivative generates an anomalous-velocity term that captures an interband contribution to the current-induced spin magnetization, the main topic of this paper.

The anomalous-velocity contribution alone, however, is insufficient to describe the full magnetoelectric response. The intraband contribution originates from a supercurrent-induced imbalance between the contributions from states with opposite momenta on each Rashba-split Fermi surface. Because the spin polarization is locked to the momentum within each helicity band, this imbalance results in a net spin magnetization \cite{ede95,yip02}. To evaluate the intraband contribution, we employ multiband Eilenberger equations formulated in terms of band-resolved quasiclassical Green’s functions \cite{nag16}. Such a band-resolved approach has been applied to superconductors with multiple Fermi surfaces \cite{nag08,uek19}. It complements the augmented Eilenberger theory with the anomalous-velocity term and enables the intraband and interband contributions to the magnetization to be evaluated separately.

In this paper, we formulate a quasiclassical theory of the Edelstein effect in superconductors that incorporates both the intraband and interband contributions. The interband contribution is described by augmented Eilenberger equations, in which the anomalous-velocity term is expressed in terms of the Berry connection associated with the momentum dependence of the Rashba eigenstates \cite{ber84,wil84,xia10}. Using this framework together with multiband Eilenberger equations, we calculate the spin magnetization as functions of temperature, distance from the surface, Rashba spin-orbit coupling strength, and supercurrent. We further investigate the nonlinear dependence of magnetization on the supercurrent and show that, within the isotropic clean $s$-wave Rashba model considered here, the nonlinear correction originates solely from the interband quantum-geometric contribution.

In Sec.~\ref{sec:ii}, we derive the gauge-covariant Wigner representation of the Gor’kov equations. In Sec.~\ref{sec:iii}, we derive the augmented Eilenberger equations in the presence of Rashba spin-orbit coupling. In Sec.~\ref{sec:iv}, we obtain expressions for the spin magnetization and current density in $s$-wave Rashba superconductors. We present numerical results for the linear and nonlinear Edelstein effects in Sec.~\ref{sec:v} and summarize our conclusions in Sec.~\ref{sec:vi}.

\section{Gor'kov equations and Wigner representation\label{sec:ii}}
\subsection{Gor'kov equations} 
The left and right Gor'kov equations, which describe an equilibrium superconductor within the mean-field approximation, are written in matrix form as \cite{uek18,kit15}
\begin{subequations}
\begin{align}
&\int d^3r_3
\left[
i\varepsilon_n\delta({\bf r}_1-{\bf r}_3)\hat{1}
-\hat{\cal H}_{\rm BdG}({\bf r}_1, {\bf r}_3)
\right] 
\hat{G}({\bf r}_3, {\bf r}_2; \varepsilon_n) \notag \\
& \ \ \ =\delta({\bf r}_1-{\bf r}_2) \hat{1}, \\
&\int d^3r_3
\hat{G}({\bf r}_1, {\bf r}_3; \varepsilon_n)
\left[
i\varepsilon_n\delta({\bf r}_3-{\bf r}_2)\hat{1} 
-\hat{\cal H}_{\rm BdG}({\bf r}_3, {\bf r}_2)
\right] \notag \\ 
& \ \ \ =\delta({\bf r}_1-{\bf r}_2)\hat{1}. 
\end{align}
\end{subequations}
Here, $\varepsilon_n=(2n+1)\pi k_{\rm B}T$ is the fermionic Matsubara energy ($n=0, \pm 1, \pm 2, \dots$), where $k_{\rm B}$ and $T$ are the Boltzmann constant and the temperature, respectively. The Green's function and the Bogoliubov-de Gennes (BdG) Hamiltonian are defined in matrix form as
\begin{subequations}
\begin{align}
\hat{G}({\bf r}_1, {\bf r}_2; \varepsilon_n) 
&\equiv
\begin{bmatrix}
\underline{G}({\bf r}_1, {\bf r}_2; \varepsilon_n) 
& \underline{F}({\bf r}_1, {\bf r}_2; \varepsilon_n) \\
-\underline{F}^*({\bf r}_1, {\bf r}_2; \varepsilon_n) 
& -\underline{G}^*({\bf r}_1, {\bf r}_2; \varepsilon_n)
\end{bmatrix}, 
\end{align}
\begin{align}
\hat{\cal H}_{\rm BdG}({\bf r}_1, {\bf r}_2) 
&\equiv
\begin{bmatrix}
\underline{\xi}({\bf r}_1, {\bf r}_2) 
& \underline{\Delta}({\bf r}_1, {\bf r}_2) \\
-\underline{\Delta}^*({\bf r}_1, {\bf r}_2) 
& -\underline{\xi}^*({\bf r}_1, {\bf r}_2)
\end{bmatrix}. 
\label{eq-H_BdG} 
\end{align}
\end{subequations}
We use a hat accent to denote matrices with Nambu-space structure and an underline accent to denote matrices in the two-dimensional spin or helicity space. In this section, hatted quantities are $4\times4$ matrices in Nambu-spin space, whereas underlined quantities are $2\times2$ matrices in spin space. Accordingly, $\hat{1}$ denotes the $4\times4$ unit matrix. The components $\underline{G}$, $\underline{F}$, $\underline{\xi}$, and $\underline{\Delta}$ satisfy the following symmetry relations:
\begin{subequations}
\begin{align}
\underline{G}({\bf r}_1, {\bf r}_2; \varepsilon_n) 
&=\underline{G}^\dagger({\bf r}_2, {\bf r}_1; -\varepsilon_n), \\
\underline{F} ({\bf r}_1, {\bf r}_2; \varepsilon_n) 
&=-\underline{F}^{\rm T} ({\bf r}_2, {\bf r}_1; -\varepsilon_n), \\
\underline{\xi}({\bf r}_1, {\bf r}_2) 
&=\underline{\xi}^\dagger({\bf r}_2, {\bf r}_1), \\
\underline{\Delta}({\bf r}_1, {\bf r}_2) 
&=-\underline{\Delta}^{\rm T}({\bf r}_2, {\bf r}_1).
\end{align}
\end{subequations}

\subsection{Gauge-covariant Wigner transform} 
It is known that the conventional Wigner transform \cite{wig32}, when applied to the Gor’kov equations, does not preserve gauge covariance with respect to the center-of-mass coordinate. Therefore, following the procedure used in Refs.~\cite{kit15,uek18}, we employ the gauge-covariant Wigner transform defined by
\begin{subequations}
\begin{align}
&\hat{G}(\varepsilon_n, {\bf p}, {\bf r}_{12}) \notag \\
&\equiv \int d^3 \bar{r}_{12}{\rm e}^{- i {\bf p}\cdot\bar{{\bf r}}_{12} / \hbar} 
\hat{\Gamma}({\bf r}_{12}, {\bf r}_1) 
\hat{G}({\bf r}_1, {\bf r}_2; \varepsilon_n) 
\hat{\Gamma} ({\bf r}_2, {\bf r}_{12}) \notag \\
&\equiv
\begin{bmatrix}
\underline{G}(\varepsilon_n, {\bf p}, {\bf r}_{12}) 
&\underline{F} (\varepsilon_n, {\bf p}, {\bf r}_{12}) \\
-\underline{F}^* (\varepsilon_n, - {\bf p}, {\bf r}_{12}) 
&-\underline{G}^* (\varepsilon_n, - {\bf p}, {\bf r}_{12})
\end{bmatrix}, 
\end{align}
where the center-of-mass and relative coordinates are defined by ${\bf r}_{12}\equiv({\bf r}_1+{\bf r}_2)/2$ and $\bar{\bf r}_{12}\equiv {\bf r}_1-{\bf r}_2$, respectively. The inverse transform is given by
\begin{align}
&\hat{G} ({\bf r}_1, {\bf r}_2; \varepsilon_n) \notag \\
&= \hat{\Gamma}( {\bf r}_1,{\bf r}_{12}) 
\int\frac{d^3 p}{(2 \pi \hbar)^3}{\rm e}^{i {\bf p} \cdot \bar{{\bf r}}_{12}/\hbar} 
\hat{G}(\varepsilon_n, {\bf p}, {\bf r}_{12}) 
\hat{\Gamma}({\bf r}_{12},{\bf r}_2). 
\end{align}
\end{subequations}
The matrix $\hat{\Gamma}$ is defined by 
\begin{align}
\hat{\Gamma} ({\bf r}_1, {\bf r}_2) &\equiv
\begin{bmatrix}
\underline{\sigma}_0{\rm e}^{i I ({\bf r}_1, {\bf r}_2)} & \underline{0} \\
\underline{0} & \underline{\sigma}_0{\rm e}^{- i I ({\bf r}_1, {\bf r}_2)}
\end{bmatrix}, 
\end{align}
where $\underline{\sigma}_0$ and $\underline{0}$ denote the $2 \times 2$ unit and zero matrices, respectively. The function $I ({\bf r}_1, {\bf r}_2)$ is the line integral defined by 
\begin{equation}
I ({\bf r}_1, {\bf r}_2) \equiv \frac{e}{\hbar} \int_{{\bf r}_2}^{{\bf r}_1} {\bf A} ({\bf s}) \cdot d {\bf s}, 
\label{I}
\end{equation}
where $e<0$ is the electron charge, ${\bf A}$ is the vector potential, and ${\bf s}$ denotes a point on the straight-line path from ${\bf r}_2$ to ${\bf r}_1$. Similarly, we transform the BdG Hamiltonian in Eq.~(\ref{eq-H_BdG}) as 
\begin{subequations}
\begin{align}
&\hat{\cal H}_{\rm BdG} ({\bf p}, {\bf r}_{12}) \notag \\
&\equiv\int d^3\bar{r}_{12}{\rm e}^{- i {\bf p} \cdot \bar{{\bf r}}_{12} / \hbar} 
\hat{\Gamma}({\bf r}_{12}, {\bf r}_1) 
\hat{\cal H}_{\rm BdG}({\bf r}_1, {\bf r}_2) 
\hat{\Gamma}({\bf r}_2, {\bf r}_{12}) \notag \\
&\equiv
\begin{bmatrix}
\underline{\xi}({\bf p}, {\bf r}_{12}) 
&\underline{\Delta}({\bf p}, {\bf r}_{12}) \\
-\underline{\Delta}^*(- {\bf p}, {\bf r}_{12}) 
&-\underline{\xi}^*(- {\bf p}, {\bf r}_{12})
\end{bmatrix}, 
\end{align}
whose inverse transform is given by
\begin{align}
&\hat{\cal H}_{\rm BdG} ({\bf r}_1, {\bf r}_2) \notag \\
&= \hat{\Gamma} ( {\bf r}_1,{\bf r}_{12}) 
\int \frac{d^3 p}{(2 \pi \hbar)^3} e^{i {\bf p} \cdot \bar{{\bf r}}_{12} / \hbar} 
\hat{\cal H}_{\rm BdG} ({\bf p}, {\bf r}_{12}) 
\hat{\Gamma} ({\bf r}_{12},{\bf r}_2). 
\end{align}
\end{subequations}
We also note that the following symmetry relations hold:
\begin{subequations}
\begin{align}
\underline{G}(\varepsilon_n, {\bf p}, {\bf r}_{12}) 
&=\underline{G}^\dagger(-\varepsilon_n, {\bf p}, {\bf r}_{12}), \\
\underline{F} (\varepsilon_n, {\bf p}, {\bf r}_{12}) 
&=-\underline{F}^{\rm T} (-\varepsilon_n, -{\bf p}, {\bf r}_{12}), \\
\underline{\xi}({\bf p}, {\bf r}_{12})
&=\underline{\xi}^\dagger({\bf p}, {\bf r}_{12}), \\
\underline{\Delta}({\bf p}, {\bf r}_{12}) 
&=-\underline{\Delta}^{\rm T}(-{\bf p}, {\bf r}_{12}). 
\end{align}
\end{subequations}

Following the same procedure as in Appendix~B of Ref.~\onlinecite{uek18}, we consider the next-to-leading-order contribution in the expansion with respect to the quasiclassical parameter $\delta\equiv\hbar/p_{\rm F}\xi_0$, where $p_{\rm F}$ is the magnitude of the Fermi momentum, $\xi_0\equiv\hbar v_{\rm F}/\Delta_0$ is the coherence length at $T=0$, $v_{\rm F}$ is the magnitude of the Fermi velocity, and $\Delta_0$ is the superconducting gap energy at $T=0$. Writing the matrix $\hat{\cal H}_{\rm BdG}({\bf p}, {\bf r})$ as 
\begin{align}
\hat{\cal H}_{\rm BdG} ({\bf p}, {\bf r})
\equiv\hat{\xi}({\bf p}, {\bf r})+\hat{\Delta}({\bf p}, {\bf r}),  
\end{align}
with $\hat{\xi}({\bf p}, {\bf r})$ and $\hat{\Delta}({\bf p}, {\bf r})$ defined by
\begin{subequations}
\begin{align}
\hat{\xi}({\bf p}, {\bf r})
&\equiv
\begin{bmatrix}
\underline{\xi}({\bf p}, {\bf r}) & \underline{0} \\
\underline{0} & - \underline{\xi}^*(-{\bf p}, {\bf r})
\end{bmatrix}, \\
\hat{\Delta}({\bf p}, {\bf r})
&\equiv
\begin{bmatrix}
\underline{0} & \underline{\Delta}({\bf p}, {\bf r}) \\
-\underline{\Delta}^*(-{\bf p}, {\bf r}) & \underline{0}
\end{bmatrix},   
\end{align}
\end{subequations}
the Gor'kov equations in the Wigner representation are given by
\begin{subequations}
\begin{align}
&\left[i\varepsilon_n\hat{1}-\hat{\xi}({\bf p}, {\bf r})-\hat{\Delta}({\bf p}, {\bf r})\right] 
*\hat{G}(\varepsilon_n, {\bf p}, {\bf r}) \notag \\
& \ \ \ -\frac{i\hbar}{8}e{\bf B}({\bf r})
\cdot
\bigg\{
\frac{\partial}{\partial{\bf p}}\hat{\xi}({\bf p}, {\bf r})\hat{\tau}_3 \notag \\
& \ \ \ \times
\frac{\partial}{\partial{\bf p}}
\left[
3\hat{G}(\varepsilon_n, {\bf p}, {\bf r})+\hat{\tau}_3\hat{G}(\varepsilon_n, {\bf p}, {\bf r})\hat{\tau}_3
\right]
\bigg\}
=\hat{1}, \\
&\hat{G}(\varepsilon_n, {\bf p}, {\bf r}) 
*\left[i\varepsilon_n \hat{1}-\hat{\xi}({\bf p}, {\bf r})-\hat{\Delta}({\bf p}, {\bf r})\right] \notag \\
& \ \ \ -\frac{i\hbar}{8}e{\bf B}({\bf r})
\cdot
\bigg\{
\frac{\partial}{\partial{\bf p}}
\left[
3\hat{G}(\varepsilon_n, {\bf p}, {\bf r})+\hat{\tau}_3\hat{G}(\varepsilon_n, {\bf p}, {\bf r})\hat{\tau}_3
\right] \notag \\
& \ \ \ \times
\frac{\partial}{\partial{\bf p}}\hat{\xi}({\bf p}, {\bf r})\hat{\tau}_3
\bigg\}
= \hat{1}, 
\end{align}
\end{subequations}
where the star product is defined by 
\begin{align}
&\hat{a}({\bf p}, {\bf r})*\hat{b}({\bf p}, {\bf r}) \notag \\
&\equiv 
{\rm e}^
{
\frac{i\hbar}{2} 
\left( 
{\bm\partial}\cdot\frac{\partial}{\partial{\bf p}'}
-\frac{\partial}{\partial{\bf p}}\cdot{\bm\partial}' 
\right)
}
\hat{a}({\bf p}, {\bf r})\hat{b}({\bf p}', {\bf r}')\bigg|_
{{\bf p}' = {\bf p}, {\bf r}'={\bf r}} \notag \\
&\approx
\hat{a}\hat{b}
+\frac{i\hbar}{2}
\left( 
{\bm\partial}\hat{a}\cdot\frac{\partial\hat{b}}{\partial{\bf p}}
-\frac{\partial\hat{a}}{\partial{\bf p}}\cdot{\bm\partial}\hat{b} 
\right),
\end{align}
with the gauge-covariant derivative ${\bm\partial}$ defined by
\begin{align}
{\bm\partial}\equiv\left\{\begin{array}{ll} 
{\bm\nabla} & {\rm on} 
\ \underline{G}, \  \underline{G}^*, 
\ \underline{\xi}, \  \underline{\xi}^* \\
\displaystyle {\bm\nabla}-i\frac{2 e {\bf A}}{\hbar} & {\rm on} 
\ \underline{F}, \ \underline{\Delta} \\
\displaystyle  {\bm\nabla}+i\frac{2 e {\bf A}}{\hbar} & {\rm on} 
\ \underline{F}^*, \ \underline{\Delta}^*
\end{array}\right. .
\end{align}
The matrix $\hat{\tau}_3$ is defined by 
\begin{align}
\hat{\tau}_3\equiv
\begin{bmatrix}
\underline{\sigma}_0 & \underline{0} \\
\underline{0} & -\underline{\sigma}_0
\end{bmatrix},  
\end{align}
and $\bf B$ is the magnetic field.

\section{Eilenberger equations with the antisymmetric spin-orbit coupling\label{sec:iii}} 
\subsection{Gor'kov equations in the Wigner representation} 
We consider superconductors with antisymmetric spin-orbit coupling. In this case, $\hat{\xi}({\bf p}, {\bf r})=\hat{\xi}_{\bf p}$ is given by
\cite{esc12,vor08}
\begin{align}
\hat{\xi}_{\bf p}= 
\xi_{\bf p} \hat{\tau}_3+{\bf g}_{\bf p}\cdot\hat{\bm\sigma}, \label{eq-xip}
\end{align}
where $\xi_{\bf p}=\xi_{-\bf p}$ is the dispersion relation measured from the chemical potential in the absence of spin-orbit coupling, ${\bf g}_{\bf p}=-{\bf g}_{-\bf p}$ is the antisymmetric spin-orbit coupling vector, and $\hat{\bm{\sigma}}$ is the vector of Pauli matrices defined by
\begin{subequations}
\begin{align}
&\hat{\bm\sigma}
\equiv
\begin{bmatrix}
\underline{\bm\sigma} & \underline{0} \\
\underline{0} & \underline{\bm\sigma}^*
\end{bmatrix}, \ \ \ 
\underline{\bm\sigma}
=
\begin{bmatrix}
\underline{\sigma}_x \\
\underline{\sigma}_y \\
\underline{\sigma}_z 
\end{bmatrix}, \\ 
&\underline{\sigma}_x
=
\begin{bmatrix}
0 & 1 \\
1 & 0
\end{bmatrix}, \ \ \ 
\underline{\sigma}_y
=
\begin{bmatrix}
0 & -i \\
i & 0
\end{bmatrix}, \ \ \ 
\underline{\sigma}_z
=
\begin{bmatrix}
1 & 0 \\
0 & -1
\end{bmatrix}.
\end{align}
\end{subequations}
The Gor'kov equations with antisymmetric spin-orbit coupling in the Wigner representation are then given by
\begin{subequations}
\begin{align}
&\left[
i\varepsilon_n\hat{1}
-\xi_{\bf p} \hat{\tau}_3-{\bf g}_{\bf p}\cdot\hat{\bm\sigma}
-\hat{\Delta}
\right] 
\hat{G} \notag \\
&+\frac{i\hbar}{2}
\left[
{\bf v}\hat{\tau}_3
+\frac{\partial}{\partial{\bf p}}
\left({\bf g}_{\bf p}\cdot\hat{\bm\sigma}\right)
\right] 
\cdot{\bm\partial}\hat{G}
-\frac{i\hbar}{2}{\bm\partial}\hat{\Delta}\cdot\frac{\partial\hat{G}}{\partial{\bf p}} \notag \\
&+\frac{i\hbar}{2}\frac{\partial\hat{\Delta}}{\partial{\bf p}}\cdot{\bm\partial}\hat{G}
-\frac{i\hbar}{8}e{\bf B}
\cdot
\left[
{\bf v}
\times\frac{\partial}{\partial{\bf p}}
\left(
3\hat{G}+\hat{\tau}_3\hat{G}\hat{\tau}_3
\right)
\right]
=\hat{1}, \label{eq-left_Gorkov} \\
&\hat{G}
\left[
i\varepsilon_n \hat{1}
-\xi_{\bf p} \hat{\tau}_3-{\bf g}_{\bf p}\cdot\hat{\bm\sigma}
-\hat{\Delta}
\right] \notag \\
&-\frac{i\hbar}{2}
{\bm\partial}\hat{G}\cdot
\left[
{\bf v}\hat{\tau}_3
+\frac{\partial}{\partial{\bf p}}
\left({\bf g}_{\bf p}\cdot\hat{\bm\sigma}\right)
\right] 
-\frac{i\hbar}{2}{\bm\partial}\hat{G}\cdot\frac{\partial\hat{\Delta}}{\partial{\bf p}} \notag \\
&+\frac{i\hbar}{2}\frac{\partial\hat{G}}{\partial{\bf p}}\cdot{\bm\partial}\hat{\Delta} 
+\frac{i\hbar}{8}e{\bf B}
\cdot
\left[
{\bf v}
\times\frac{\partial}{\partial{\bf p}}
\left(
3\hat{G}+\hat{\tau}_3\hat{G}\hat{\tau}_3
\right)
\right]
=\hat{1}. \label{eq-right_Gorkov} 
\end{align} 
\end{subequations}
Here, ${\bf v}$ is the ordinary velocity defined by ${\bf v}\equiv\partial\xi_{\bf p}/\partial{\bf p}$. In the Lorentz-force term, we retain only this ordinary velocity because the Lorentz-force term is already of first order in the quasiclassical parameter. Including the anomalous velocity arising from $\partial({\bf g}_{\bf p}\cdot\hat{\bm\sigma})/\partial{\bf p}$ therefore produces a second-order contribution, and it can be neglected.

\subsection{Augmented Eilenberger equations}
We first derive the main part of the augmented Eilenberger equations. For this purpose, we multiply both sides of Eq.~(\ref{eq-right_Gorkov}) by $\hat{\tau}_3$ and subtract the resulting equation from Eq.~(\ref{eq-left_Gorkov}). The resulting equation is rewritten in terms of $\hat{\tau}_3\hat{G}$. We then obtain
\begin{align}
&\left[ 
i\varepsilon_n\hat{\tau}_3
-{\bf g}_{\bf p}\cdot\hat{\bm\sigma}\hat{\tau}_3
-\hat{\Delta}\hat{\tau}_3, 
\hat{\tau}_3\hat{G} 
\right]
+i\hbar{\bf v}\cdot{\bm\partial}\hat{\tau}_3\hat{G} \notag \\
& \ \ \ -\frac{i\hbar}{2}e{\bf B}
\cdot
\left(
{\bf v}\times\frac{\partial}{\partial{\bf p}}
\right)
\left\{
\hat{\tau}_3, \hat{\tau}_3\hat{G}
\right\} \notag \\
& \ \ \ 
+\frac{i\hbar}{2}
\frac{\partial}{\partial{\bf p}}
\left({\bf g}_{\bf p}\cdot\hat{\bm\sigma}\hat{\tau}_3\right)
\cdot
{\bm\partial}\hat{\tau}_3\hat{G}
+\frac{i\hbar}{2}
{\bm\partial}\hat{\tau}_3\hat{G}
\cdot
\frac{\partial}{\partial{\bf p}}
\left({\bf g}_{\bf p}\cdot\hat{\bm\sigma}\hat{\tau}_3\right) \notag \\
& \ \ \ 
-\frac{i\hbar}{2}{\bm\partial}\hat{\Delta}\hat{\tau}_3
\cdot\frac{\partial\hat{\tau}_3\hat{G}}{\partial{\bf p}}
+\frac{i\hbar}{2}
\frac{\partial\hat{\Delta}\hat{\tau}_3}{\partial{\bf p}}
\cdot{\bm\partial}\hat{\tau}_3\hat{G} \notag \\
& \ \ \ 
+\frac{i\hbar}{2}{\bm\partial}\hat{\tau}_3\hat{G}
\cdot\frac{\partial\hat{\Delta}\hat{\tau}_3}{\partial{\bf p}}
-\frac{i\hbar}{2}
\frac{\partial\hat{\tau}_3\hat{G}}{\partial{\bf p}}
\cdot{\bm\partial}\hat{\Delta}\hat{\tau}_3 
=\hat{0}, 
\label{eq-left-right_Gorkov}
\end{align}
where $\hat{0}$ denotes the $4\times 4$ zero matrix. The commutator $[\hat{a},\hat{b}]$ and anticommutator $\{\hat{a},\hat{b}\}$ are defined by $[\hat{a},\hat{b}]=\hat{a}\hat{b}-\hat{b}\hat{a}$ and $\{\hat{a},\hat{b}\}=\hat{a}\hat{b}+\hat{b}\hat{a}$, respectively.

We now introduce the quasiclassical Green's function
\begin{align}
&\hat{g}(\varepsilon_n, {\bf p}_{\rm F}, {\bf r}) 
\equiv 
{\rm P} \int_{- \infty}^\infty \frac{d \xi_{\bf p}}{\pi} 
i\hat{\tau}_3\hat{G}(\varepsilon_n, {\bf p}, {\bf r}) \notag \\
&\equiv
\begin{bmatrix}
\underline{g}(\varepsilon_n, {\bf p}_{\rm F}, {\bf r}) 
&-i\underline{f}(\varepsilon_n, {\bf p}_{\rm F}, {\bf r}) \\
-i\underline{f}^*(\varepsilon_n, -{\bf p}_{\rm F}, {\bf r}) 
&-\underline{g}^*(\varepsilon_n, -{\bf p}_{\rm F}, {\bf r})
\end{bmatrix}, 
\end{align}
where ${\rm P}$ denotes the principal value. It follows that the components $\underline{g}$ and $\underline{f}$ satisfy
\begin{subequations}
\begin{align}
\underline{g}(\varepsilon_n, {\bf p}_{\rm F}, {\bf r}) 
&=-\underline{g}^\dagger(-\varepsilon_n, {\bf p}_{\rm F}, {\bf r}), \\
\underline{f}(\varepsilon_n, {\bf p}_{\rm F}, {\bf r}) 
&=-\underline{f}^{\rm T}(-\varepsilon_n, -{\bf p}_{\rm F}, {\bf r}). 
\end{align}
\end{subequations}
We also write $\underline{g}$ and $\underline{f}$ as
\begin{align}
\underline{g}\equiv
\begin{bmatrix}
g_{\uparrow\uparrow}
&g_{\uparrow\downarrow} \\
g_{\downarrow\uparrow} 
&g_{\downarrow\downarrow}
\end{bmatrix}, \ \ \ 
\underline{f}\equiv
\begin{bmatrix}
f_{\uparrow\uparrow}
&f_{\uparrow\downarrow} \\
f_{\downarrow\uparrow} 
&f_{\downarrow\downarrow}
\end{bmatrix}. 
\end{align}

To derive the equation for $\hat{g}$ from Eq.~(\ref{eq-left-right_Gorkov}), we decompose the momentum derivative as $\partial/\partial{\bf p}=\partial/\partial{\bf p}_\parallel+{\bf v}(\partial/\partial\xi_{\bf p})$, where ${\bf p}_\parallel$ denotes the component tangent to the energy surface $\xi_{\bf p}={\rm const}$. We then set ${\bf p}={\bf p}_{\rm F}$ except in the argument of $\hat{G}$, integrate Eq.~(\ref{eq-left-right_Gorkov}) over $-\varepsilon_{\rm c}\le\xi_{\bf p}\le\varepsilon_{\rm c}$, use ${\bf v}\times(\partial/\partial{\bf p}_\parallel)={\bf v}\times(\partial/\partial{\bf p})$ and
\begin{align}
{\rm P}\int_{-\infty}^\infty d\xi_{\bf p}
\frac{\partial}{\partial\xi_{\bf p}}
\hat{G}(\varepsilon_n, {\bf p}, {\bf r})
=\hat{0},
\end{align}
and finally take the limit $\varepsilon_{\rm c}\to\infty$. We thereby obtain the main part of the augmented Eilenberger equations as
\begin{align}
&\left[ 
i\varepsilon_n\hat{\tau}_3
-{\bf g}_{{\bf p}_{\rm F}}\cdot\hat{\bm\sigma}\hat{\tau}_3
-\hat{\Delta}\hat{\tau}_3, 
\hat{g} 
\right]
+i\hbar{\bf v}_{\rm F}\cdot{\bm\partial}\hat{g} \notag \\
& \ \ \ 
+\frac{i\hbar}{2}e({\bf v}_{\rm F}\times{\bf B})
\cdot
\frac{\partial}{\partial{\bf p}_{\rm F}}
\{\hat{\tau}_3, \hat{g}\} 
\notag \\
& \ \ \ 
+\frac{i\hbar}{2}
\frac{\partial}{\partial{\bf p}_{\rm F}}
\left({\bf g}_{{\bf p}_{\rm F}}\cdot\hat{\bm\sigma}\hat{\tau}_3\right)
\cdot
{\bm\partial}\hat{g}
+\frac{i\hbar}{2}
{\bm\partial}\hat{g}
\cdot
\frac{\partial}{\partial{\bf p}_{\rm F}}
\left({\bf g}_{{\bf p}_{\rm F}}\cdot\hat{\bm\sigma}\hat{\tau}_3\right) \notag \\
& \ \ \ 
-\frac{i\hbar}{2}{\bm\partial}\hat{\Delta}\hat{\tau}_3
\cdot\frac{\partial\hat{g}}{\partial{\bf p}_{\rm F}}
+\frac{i\hbar}{2}
\frac{\partial\hat{\Delta}\hat{\tau}_3}{\partial{\bf p}_{\rm F}}
\cdot{\bm\partial}\hat{g} \notag \\ 
& \ \ \ 
+\frac{i\hbar}{2}{\bm\partial}\hat{g}
\cdot\frac{\partial\hat{\Delta}\hat{\tau}_3}{\partial{\bf p}_{\rm F}}
-\frac{i\hbar}{2}
\frac{\partial\hat{g}}{\partial{\bf p}_{\rm F}}
\cdot{\bm\partial}\hat{\Delta}\hat{\tau}_3 
=\hat{0}, 
\label{eq-augmented_Eilenberger}
\end{align}
where ${\bf v}_{\rm F}$ is defined by ${\bf v}_{\rm F}\equiv\partial\xi_{\bf p}/\partial{\bf p}|_{{\bf p}={\bf p}_{\rm F}}$.

\subsection{Rashba superconductors}
We consider superconductors with Rashba spin-orbit coupling given by
\cite{hay06,esc12,vor08}
\begin{align}
{\bf g}_{{\bf p}_{\rm F}} 
= 
\begin{bmatrix}
-\alpha\sin\varphi_{\bf p} \\
\alpha\cos\varphi_{\bf p} \\
0
\end{bmatrix}, 
\end{align}
where $\alpha$ is the strength of the spin-orbit coupling and $\varphi_{\bf p}$ is the azimuthal angle of the momentum vector. To diagonalize ${\bf g}_{{\bf p}_{\rm F}}\cdot\hat{\bm\sigma}\hat{\tau}_3$ in Eq.~(\ref{eq-augmented_Eilenberger}), we introduce the unitary matrix \cite{esc12,vor08}
\begin{align}
\hat{U}_{{\bf p}_{\rm F}}
\equiv 
\begin{bmatrix}
\underline{U}_{{\bf p}_{\rm F}} & \underline{0} \\
\underline{0} & \underline{U}_{-{\bf p}_{\rm F}}^*
\end{bmatrix}, \ \ \ 
\underline{U}_{{\bf p}_{\rm F}} 
\equiv
\frac{1}{\sqrt{2}}
\begin{bmatrix}
-i{\rm e}^{-i\varphi_{\bf p}} &1 \\
1 & -i{\rm e}^{i\varphi_{\bf p}} 
\end{bmatrix}. 
\label{eq-U_p}
\end{align}
Then, we obtain
\begin{subequations}
\begin{align}
&\hat{U}_{{\bf p}_{\rm F}}^\dagger
{\bf g}_{{\bf p}_{\rm F}}\cdot\hat{\bm\sigma}\hat{\tau}_3
\hat{U}_{{\bf p}_{\rm F}}
=\alpha\hat{\sigma}_z, 
\label{eq-band_energy} \\
&\hat{U}_{{\bf p}_{\rm F}}^\dagger 
\left[
\frac{\partial}{\partial{\bf p}_{\rm F}}
\left(
{\bf g}_{{\bf p}_{\rm F}}\cdot\hat{\bm\sigma}\hat{\tau}_3
\right)
\right]
\hat{U}_{{\bf p}_{\rm F}}
=
\left[
\hat{U}_{{\bf p}_{\rm F}}^\dagger
\frac{\partial\hat{U}_{{\bf p}_{\rm F}}}{\partial{\bf p}_{\rm F}}, 
\alpha\hat{\sigma}_z
\right], 
\label{eq-anomalous_velocity}
\end{align}
\end{subequations}
where $\hat{\sigma}_z$ is defined by
\begin{align}
\hat{\sigma}_z
\equiv 
\begin{bmatrix}
\underline{\sigma}_z & \underline{0} \\
\underline{0} & \underline{\sigma}_z^*
\end{bmatrix}. 
\end{align}
Equation~(\ref{eq-anomalous_velocity}) represents the anomalous velocity, which contributes to the interband Green's functions. We also calculate $\hat{U}_{{\bf p}_{\rm F}}^\dagger(\partial\hat{g}/\partial{\bf p}_{\rm F})\hat{U}_{{\bf p}_{\rm F}}$ and $\hat{U}_{{\bf p}_{\rm F}}^\dagger[\partial(\hat{\Delta}\hat{\tau}_3)/\partial{\bf p}_{\rm F}]\hat{U}_{{\bf p}_{\rm F}}$
as
\begin{subequations}
\begin{align}
&\hat{U}_{{\bf p}_{\rm F}}^\dagger
\frac{\partial\hat{g}}{\partial{\bf p}_{\rm F}}
\hat{U}_{{\bf p}_{\rm F}}
=
\frac{\partial\hat{g}'}{\partial{\bf p}_{\rm F}}
+
\left[
\hat{U}_{{\bf p}_{\rm F}}^\dagger
\frac{\partial\hat{U}_{{\bf p}_{\rm F}}}{\partial{\bf p}_{\rm F}}, 
\hat{g}'
\right], \\
&\hat{U}_{{\bf p}_{\rm F}}^\dagger
\frac{\partial\hat{\Delta}\hat{\tau}_3}{\partial{\bf p}_{\rm F}}
\hat{U}_{{\bf p}_{\rm F}}
=
\frac{\partial\hat{\Delta}'\hat{\tau}_3}{\partial{\bf p}_{\rm F}}
+
\left[
\hat{U}_{{\bf p}_{\rm F}}^\dagger
\frac{\partial\hat{U}_{{\bf p}_{\rm F}}}{\partial{\bf p}_{\rm F}}, 
\hat{\Delta}'\hat{\tau}_3
\right],
\end{align}\label{eq-derivative-g-Delta}%
\end{subequations}
with
$\hat{g}'\equiv\hat{U}_{{\bf p}_{\rm F}}^\dagger\hat{g}\hat{U}_{{\bf p}_{\rm F}}$ and $\hat{\Delta}'\equiv\hat{U}_{{\bf p}_{\rm F}}^\dagger\hat{\Delta}\hat{U}_{{\bf p}_{\rm F}}$. The matrix $\hat{\cal A}_{{\bf p}_{\rm F}}\equiv i\hat{U}_{{\bf p}_{\rm F}}^\dagger(\partial\hat{U}_{{\bf p}_{\rm F}}/\partial{\bf p}_{\rm F})$ is the non-Abelian Berry connection of the helicity basis evaluated on the Fermi surface \cite{ber84,wil84,xia10}. The corresponding connection terms in Eqs.~(\ref{eq-anomalous_velocity}) and (\ref{eq-derivative-g-Delta}) originate from the momentum dependence of the Rashba helicity eigenstates, giving rise to quantum geometric properties in Rashba systems. Using these results, we obtain the main part of the augmented Eilenberger equations for Rashba superconductors as
\begin{align}
&\left[ 
i\varepsilon_n\hat{\tau}_3
-\alpha\hat{\sigma}_z
-\hat{\Delta}'\hat{\tau}_3, 
\hat{g}' 
\right] \notag \\
& \ \ \ +\frac{i\hbar}{2}
\left\{
{\bf v}_{\rm F}\hat{1}-i\left[\hat{\cal A}_{{\bf p}_{\rm F}}, \alpha\hat{\sigma}_z\right], 
{\bm\partial}\hat{g}' 
\right\} \notag \\
& \ \ \ 
+\frac{i\hbar}{2}e({\bf v}_{\rm F}\times{\bf B})
\cdot
\left\{
\hat{\tau}_3, 
\frac{\partial\hat{g}'}{\partial{\bf p}_{\rm F}}
-i\left[\hat{\cal A}_{{\bf p}_{\rm F}}, \hat{g}'\right]
\right\}
\notag \\
& \ \ \ 
-\frac{i\hbar}{2}
\bigg\{
{\bm\partial}\hat{\Delta}'\hat{\tau}_3, 
\frac{\partial\hat{g}'}{\partial{\bf p}_{\rm F}} 
-i\left[\hat{\cal A}_{{\bf p}_{\rm F}}, \hat{g}'\right]
\bigg\} \notag \\
& \ \ \ 
+\frac{i\hbar}{2}
\Bigg\{
\frac{\partial\hat{\Delta}'\hat{\tau}_3}{\partial{\bf p}_{\rm F}} 
-i\left[\hat{\cal A}_{{\bf p}_{\rm F}}, \hat{\Delta}'\hat{\tau}_3\right], 
{\bm\partial}\hat{g}' 
\Bigg\} 
=\hat{0}. 
\label{eq-augmented_Eilenberger_Rashba}
\end{align}

\section{Edelstein effect in s-wave Rashba superconductors\label{sec:iv}}
\subsection{Augmented Eilenberger equations with the anomalous velocity}
For simplicity, we restrict ourselves to $s$-wave superconductors and neglect the Lorentz-force and pair-potential-gradient terms, which will be justified later. Then, the main part of the augmented Eilenberger equations including the anomalous velocity is given by
\begin{align}
&\left[ 
i\varepsilon_n\hat{\tau}_3
-\alpha\hat{\sigma}_z
-\hat{\Delta}'\hat{\tau}_3, 
\hat{g}' 
\right]
+\frac{i\hbar}{2}
\left\{
\hat{\bf v}_{\rm F}', 
{\bm\partial}\hat{g}' 
\right\}
=\hat{0}, 
\label{eq-augmented_Eilenberger_anomalous_velocity}
\end{align}
where $\hat{\bf v}_{\rm F}'$ is the velocity matrix defined by
\begin{align}
\hat{\bf v}_{\rm F}'
&\equiv
{\bf v}_{\rm F}\hat{1}
-i\left[\hat{\cal A}_{{\bf p}_{\rm F}}, \alpha\hat{\sigma}_z\right] \notag \\
&=
{\bf v}_{\rm F}\hat{1}
-\alpha\frac{\partial\varphi_{\bf p}}{\partial{\bf p}_{\rm F}}
(\cos\varphi_{\bf p}\hat{\sigma}_x
-\sin\varphi_{\bf p}\hat{\sigma}_y)\hat{\tau}_3, 
\label{eq-effective_velocity_Rashba}
\end{align}
with $\hat{\sigma}_x$ and $\hat{\sigma}_y$ defined by
\begin{align}
\hat{\sigma}_x
\equiv 
\begin{bmatrix}
\underline{\sigma}_x & \underline{0} \\
\underline{0} & \underline{\sigma}_x^*
\end{bmatrix}, \ \ \
\hat{\sigma}_y
\equiv 
\begin{bmatrix}
\underline{\sigma}_y & \underline{0} \\
\underline{0} & \underline{\sigma}_y^*
\end{bmatrix}. 
\end{align}
The second term in $\hat{\bf v}_{\rm F}'$ is the anomalous velocity and has interband off-diagonal elements in the helicity basis. The Green's function $\hat{g}'=\hat{g}'(\varepsilon_n,{\bf p}_{\rm F},{\bf r})$ is expressed as
\begin{align}
\hat{g}'
\equiv
\begin{bmatrix}
\underline{g}' 
&-i\underline{f}' \\
-i\underline{\bar{f}}'
&-\underline{\bar{g}}'
\end{bmatrix}, 
\underline{g}' 
\equiv
\begin{bmatrix}
g_{++} 
&g_{+-} \\
g_{-+}
&g_{--}
\end{bmatrix}, 
\underline{f}' 
\equiv
\begin{bmatrix}
f_{++} 
&f_{+-} \\
f_{-+}
&f_{--}
\end{bmatrix}.
\end{align}
For spin-singlet $s$-wave pairing, the pair potential in spin space is given by
\begin{align}
\underline{\Delta}({\bf r})
=
\Delta({\bf r})i\underline{\sigma}_y .
\end{align}
In the helicity basis, the pair potential
$\hat{\Delta}'=\hat{\Delta}'({\bf p}_{\rm F},{\bf r})$ is expressed as
\begin{subequations}
\begin{align}
\hat{\Delta}'
&\equiv
\begin{bmatrix}
\underline{0} 
&\underline{\Delta}' \\
-\underline{\bar{\Delta}}'
&\underline{0}
\end{bmatrix}, \\
\underline{\Delta}' 
&\equiv
\begin{bmatrix}
\Delta_{++} 
&\Delta_{+-} \\
\Delta_{-+}
&\Delta_{--}
\end{bmatrix}
=
\begin{bmatrix}
i\Delta({\bf r}){\rm e}^{i\varphi_{\bf p}}
&0 \\
0
&-i\Delta({\bf r}){\rm e}^{-i\varphi_{\bf p}}
\end{bmatrix}.
\label{eq-Delta_s-wave_Rashba}
\end{align}
\end{subequations}
Here, we define $\bar{X}(\varepsilon_n,{\bf p}_{\rm F},{\bf r})\equiv X^*(\varepsilon_n,-{\bf p}_{\rm F},{\bf r})$ for arbitrary $X$.

\subsection{London limit and expansion in powers of the quasiclassical parameter}
We consider a superconductor in the London limit and express the pair potential and anomalous Green's functions as
\begin{subequations}
\begin{align}
\Delta({\bf r})&=|\Delta({\bf r})|{\rm e}^{i\varphi({\bf r})}, \\
\underline{f}'(\varepsilon_n,{\bf p}_{\rm F},{\bf r})
&=
\underline{\tilde{f}}'(\varepsilon_n,{\bf p}_{\rm F},{\bf r})
{\rm e}^{i\varphi({\bf r})}.
\end{align}
\end{subequations}
We also neglect the spatial derivatives of $\underline{g}'$ and $\underline{\tilde{f}}'$. Then, the augmented Eilenberger equations become
\begin{align}
&\left[ 
i\varepsilon_n\hat{\tau}_3
-\alpha\hat{\sigma}_z
-\hat{\Delta}'\hat{\tau}_3, 
\hat{g}' 
\right]
-\frac{1}{2}
\left\{
\hat{\bf v}_{\rm F}'\cdot{\bf p}_{\rm s}, 
\left[\hat{\tau}_3,\hat{g}'\right] 
\right\}
=\hat{0}, 
\label{eq-augmented_Eilenberger_London}
\end{align}
where ${\bf p}_{\rm s}$ is the superfluid momentum defined by
\begin{align}
{\bf p}_{\rm s}
\equiv
\frac{\hbar}{2}
\left(
{\bm\nabla}\varphi-\frac{2e}{\hbar}{\bf A}
\right).
\end{align}

We expand $\hat{g}'$ formally in powers of the quasiclassical parameter $\delta$ as $\hat{g}'=\hat{g}'{}^{(0)}+\hat{g}'{}^{(1)}+\cdots$. Then, $\hat{g}'{}^{(0)}$ is the solution of the standard Eilenberger equations,
\begin{align}
&\left[ 
(i\varepsilon_n-{\bf v}_{\rm F}\cdot{\bf p}_{\rm s})\hat{\tau}_3
-\alpha\hat{\sigma}_z
-\hat{\Delta}'\hat{\tau}_3, 
\hat{g}'{}^{(0)} 
\right]
=\hat{0}, 
\ \
\hat{g}'{}^{(0)2}=\hat{1}.
\label{eq-Eilenberger_zeroth}
\end{align}
The equation for the first-order correction in $\delta$ is given by
\begin{align}
&\left[ 
(i\varepsilon_n-{\bf v}_{\rm F}\cdot{\bf p}_{\rm s})\hat{\tau}_3
-\alpha\hat{\sigma}_z
-\hat{\Delta}'\hat{\tau}_3, 
\hat{g}'{}^{(1)} 
\right] \notag \\
& \ \ \ 
-\frac{1}{2}
\left\{
(\hat{\bf v}_{\rm F}'-{\bf v}_{\rm F}\hat{1})\cdot{\bf p}_{\rm s}, 
\left[\hat{\tau}_3,\hat{g}'{}^{(0)}\right] 
\right\}
=\hat{0}.
\label{eq-augmented_Eilenberger_first}
\end{align}

\subsection{Current density and interband magnetization}
We calculate the current density and magnetization using the following expressions derived in Appendix~\ref{AppA} and \ref{AppB}:
\begin{subequations}
\begin{align}
&{\bf j}
=-i\pi k_{\rm B}TeN(0) \notag \\
&\times
\sum_{n=-\infty}^{\infty}
\left\langle
{\rm Tr}
\left\{
\left[
{\bf v}_{\rm F}\underline{\sigma}_0
+
\frac{\partial}{\partial{\bf p}_{\rm F}}
\left(
{\bf g}_{{\bf p}_{\rm F}}\cdot\underline{\bm\sigma}
\right)
\right]
\underline{g}
\right\}
\right\rangle_{\rm F},
\label{eq-j} \\
&{\bf M}
=i\pi k_{\rm B}T\mu_{\rm B}N(0)
\sum_{n=-\infty}^{\infty}
\left\langle
{\rm Tr}
\underline{\bm\sigma}\,\underline{g}
\right\rangle_{\rm F}.
\label{eq-M}
\end{align}
\label{eq-j_M}%
\end{subequations}
Here, $N(0)$ is the normal-state density of states at the Fermi level, $\langle\cdots\rangle_{\rm F}$ denotes the Fermi-surface average normalized as $\langle1\rangle_{\rm F}=1$, $\mu_{\rm B}\equiv |e|\hbar/2m$ is the Bohr magneton, and $m$ is the electron mass. We note that Eq.~(\ref{eq-j_M}) is written in spin space.

To this end, we first solve the standard Eilenberger equations, Eq.~(\ref{eq-Eilenberger_zeroth}). Since both the Rashba spin-orbit coupling term and the transformed pair potential $\hat{\Delta}'$ in Eq.~(\ref{eq-Delta_s-wave_Rashba}) are diagonal in helicity space, Eq.~(\ref{eq-Eilenberger_zeroth}) decouples into separate equations for the two helicity bands. Accordingly, the zeroth-order quasiclassical Green's function $\hat{g}'^{(0)}$ is diagonal in helicity space and obtained as
\begin{subequations}
\begin{align}
&\hat{g}'{}^{(0)}
\equiv
\begin{bmatrix}
\underline{g}'{}^{(0)} 
&-i\underline{f}'{}^{(0)} \\
-i\underline{\bar{f}}'{}^{(0)}
&-\underline{\bar{g}}'{}^{(0)}
\end{bmatrix}, \\ 
&\underline{g}'{}^{(0)} 
\equiv
\begin{bmatrix}
g_{++}^{(0)} 
&0 \\
0
&g_{--}^{(0)}
\end{bmatrix}, \ \ \ 
\underline{f}'{}^{(0)} 
\equiv
\begin{bmatrix}
f_{++}^{(0)} 
&0 \\
0
&f_{--}^{(0)}
\end{bmatrix}, 
\end{align}
and
\begin{align}
&g_{++}^{(0)}
=g_{--}^{(0)}
=
\frac{\tilde{\varepsilon}_n}
{\sqrt{\tilde{\varepsilon}_n^2+|\Delta|^2}}, \\
&f_{++}^{(0)}
=
-f_{--}^{(0)}{\rm e}^{2i\varphi_{\bf p}}
=
\frac{i\Delta{\rm e}^{i\varphi_{\bf p}}}
{\sqrt{\tilde{\varepsilon}_n^2+|\Delta|^2}},
\end{align}
\end{subequations}
where $\tilde{\varepsilon}_n\equiv\varepsilon_n+i{\bf v}_{\rm F}\cdot{\bf p}_{\rm s}$. Using the unitary matrix Eq.~(\ref{eq-U_p}), we also obtain the Green's functions in spin space to zeroth order in $\delta$ as
\begin{subequations}
\begin{align}
&g_{\uparrow\uparrow}^{(0)}
=g_{\downarrow\downarrow}^{(0)}
=
\frac{\tilde{\varepsilon}_n}
{\sqrt{\tilde{\varepsilon}_n^2+|\Delta|^2}}, \\
&f_{\uparrow\downarrow}^{(0)}
=
-f_{\downarrow\uparrow}^{(0)}
=
\frac{\Delta}
{\sqrt{\tilde{\varepsilon}_n^2+|\Delta|^2}}.
\end{align}
\label{eq-g^(0)_f^(0)_spin}%
\end{subequations}
We note that, in the present spin-singlet $s$-wave state, the Lorentz-force and pair-potential-gradient terms do not generate spin polarization to first order in the quasiclassical parameter $\delta$. This can be verified by perturbatively substituting Eq.~(\ref{eq-g^(0)_f^(0)_spin}) into those terms in Eq.~(\ref{eq-augmented_Eilenberger}).

We next solve the first-order correction to the standard Eilenberger equations, Eq.~(\ref{eq-augmented_Eilenberger_first}). The diagonal components $g_{++}^{(1)}$, $g_{--}^{(1)}$, $f_{++}^{(1)}$, $f_{--}^{(1)}$, and the corresponding barred functions satisfy a closed set of homogeneous linear equations because the first-order driving terms are off-diagonal in helicity space. We therefore choose the trivial solution, $g_{++}^{(1)}=g_{--}^{(1)}=f_{++}^{(1)}=f_{--}^{(1)}=0$. Thus, $\hat{g}'{}^{(1)}$ is expressed as
\begin{subequations}
\begin{align}
&\hat{g}'{}^{(1)}
\equiv
\begin{bmatrix}
\underline{g}'{}^{(1)} 
&-i\underline{f}'{}^{(1)} \\
-i\underline{\bar{f}}'{}^{(1)}
&-\underline{\bar{g}}'{}^{(1)}
\end{bmatrix}, \\ 
&\underline{g}'{}^{(1)} 
\equiv
\begin{bmatrix}
0
&g_{+-}^{(1)}  \\
g_{-+}^{(1)} 
&0
\end{bmatrix}, \ \ \ 
\underline{f}'{}^{(1)} 
\equiv
\begin{bmatrix}
0
&f_{+-}^{(1)}  \\
f_{-+}^{(1)} 
&0
\end{bmatrix}. 
\end{align}
\end{subequations}
Equations for $g_{+-}^{(1)}$, $g_{-+}^{(1)}$, $f_{+-}^{(1)}$, and $f_{-+}^{(1)}$ are given by
\begin{subequations}
\begin{align}
&2\alpha g_{+-}^{(1)}
+i(\Delta_{++}\bar f_{+-}^{(1)} - f_{+-}^{(1)}\bar\Delta_{--})=0, \\
&2\alpha g_{-+}^{(1)}
-i(\Delta_{--}\bar f_{-+}^{(1)} - f_{-+}^{(1)}\bar\Delta_{++})=0, \\
&2(\tilde\varepsilon_n+i\alpha)f_{+-}^{(1)}
-(\Delta_{++}\bar g_{+-}^{(1)} + g_{+-}^{(1)}\Delta_{--}) \notag \\
&=-i\alpha\frac{\partial\varphi_{\bf p}}{\partial{\bf p}_{\rm F}}\cdot{\bf p}_{\rm s}
(f_{++}^{(0)}e^{-i\varphi_{\bf p}} - f_{--}^{(0)}e^{i\varphi_{\bf p}}), \\
&2(\tilde\varepsilon_n-i\alpha)f_{-+}^{(1)}
-(\Delta_{--}\bar g_{-+}^{(1)} + g_{-+}^{(1)}\Delta_{++}) \notag \\
&=i\alpha\frac{\partial\varphi_{\bf p}}{\partial{\bf p}_{\rm F}}\cdot{\bf p}_{\rm s}
(f_{++}^{(0)}e^{-i\varphi_{\bf p}} - f_{--}^{(0)}e^{i\varphi_{\bf p}}).
\end{align}
\end{subequations}
Solving these simultaneous linear equations, we obtain
\begin{subequations}
\begin{align}
g_{+-}^{(1)} 
&= 
-\frac{i\alpha|\Delta|^2 e^{i\varphi_{\bf p}}}
{(\tilde\varepsilon_n^2+\alpha^2+|\Delta|^2)
\sqrt{\tilde\varepsilon_n^2+|\Delta|^2}}
\frac{\partial\varphi_{\bf p}}{\partial{\bf p}_{\rm F}}\cdot{\bf p}_{\rm s}, \\
g_{-+}^{(1)} 
&= 
-\frac{i\alpha|\Delta|^2 e^{-i\varphi_{\bf p}}}
{(\tilde\varepsilon_n^2+\alpha^2+|\Delta|^2)
\sqrt{\tilde\varepsilon_n^2+|\Delta|^2}}
\frac{\partial\varphi_{\bf p}}{\partial{\bf p}_{\rm F}}\cdot{\bf p}_{\rm s}, \\
f_{+-}^{(1)} 
&= 
\frac{\alpha(\tilde\varepsilon_n - i\alpha)\,\Delta}
{(\tilde\varepsilon_n^2+\alpha^2+|\Delta|^2)
\sqrt{\tilde\varepsilon_n^2+|\Delta|^2}}
\frac{\partial\varphi_{\bf p}}{\partial{\bf p}_{\rm F}}\cdot{\bf p}_{\rm s}, \\
f_{-+}^{(1)} 
&= 
-\frac{\alpha(\tilde\varepsilon_n + i\alpha)\,\Delta}
{(\tilde\varepsilon_n^2+\alpha^2+|\Delta|^2)
\sqrt{\tilde\varepsilon_n^2+|\Delta|^2}}
\frac{\partial\varphi_{\bf p}}{\partial{\bf p}_{\rm F}}\cdot{\bf p}_{\rm s}.
\end{align}
\end{subequations}
Using $g_{++}^{(0)}$, $g_{+-}^{(1)}$, $g_{-+}^{(1)}$, $g_{--}^{(0)}$, and Eq.~(\ref{eq-U_p}), we obtain the quasiclassical Green functions in the spin basis $g_{\uparrow\uparrow}$, $g_{\uparrow\downarrow}$, $g_{\downarrow\uparrow}$, and $g_{\downarrow\downarrow}$ as
\begin{subequations}
\begin{align}
g_{\uparrow\uparrow}
&=g_{\downarrow\downarrow}
=g_{++}^{(0)}
=g_{--}^{(0)}, \\
g_{\uparrow\downarrow}
&=g_{\downarrow\uparrow}{\rm e}^{-2i\varphi_{\bf p}}
=g_{-+}^{(1)}
=g_{+-}^{(1)}{\rm e}^{-2i\varphi_{\bf p}}. 
\end{align}
\label{eq-g_spin}%
\end{subequations}
Substituting Eq.~(\ref{eq-g_spin}) into Eq.~(\ref{eq-j_M}), we finally obtain the current density and interband contribution to the magnetization as
\begin{subequations}
\begin{align}
&{\bf j}
=
-2i\pi k_{\rm B}TeN(0)
\sum_{n=-\infty}^\infty
\left\langle
{\bf v}_{\rm F}
\frac{\tilde{\varepsilon}_n}
{\sqrt{\tilde{\varepsilon}_n^2+|\Delta|^2}}
\right\rangle_{\rm F}, 
\label{eq-j-nonlinear} \\
&{\bf M}^{\rm inter} 
=
-2\pi k_{\rm B}T\mu_{\rm B}N(0)\sum_{n=-\infty}^{\infty} \notag \\
&\times\left\langle
\frac{|\Delta|^2}
{(\tilde\varepsilon_n^2+\alpha^2+|\Delta|^2)
\sqrt{\tilde\varepsilon_n^2+|\Delta|^2}}
{\bf p}_{\rm s}\cdot\frac{\partial}{\partial{\bf p}_{\rm F}}{\bf g}_{{\bf p}_{\rm F}}
\right\rangle_{\rm F}.
\label{eq-M_inter-nonlinear} 
\end{align}
\end{subequations}
The superconducting gap $|\Delta|$ is determined by the gap equation
\begin{align}
\ln\frac{T}{T_{\rm c}}
+\pi k_{\rm B}T
\sum_{n=-\infty}^{\infty}
\left(
\frac{1}{|\varepsilon_n|}
-
\left\langle
\frac{1}
{\sqrt{\tilde{\varepsilon}_n^2+|\Delta|^2}}
\right\rangle_{\rm F}
\right)
=0,
\label{eq-gap-nonlinear}
\end{align}
where $T_{\rm c}$ is the transition temperature. We note that the interband anomalous components $f_{+-}^{(1)}$ and $f_{-+}^{(1)}$ do not induce a first-order correction to the $s$-wave spin-singlet pair potential, because they cancel in the spin-singlet combination after transforming back to the spin basis.

\subsection{Multiband Eilenberger equations and intraband magnetization} 
We have calculated the interband contribution to magnetization using the augmented Eilenberger equations including the anomalous velocity. The Doppler shift also produces an imbalance in the occupation of quasiparticle states with opposite momenta within each helicity band. Because these states carry opposite spin polarizations, this imbalance gives rise to an intraband contribution to magnetization. Here, we calculate the intraband contribution using the multiband Eilenberger equations \cite{nag08,nag16,uek19}.

We evaluate the two contributions using complementary quasiclassical formulations that retain different sectors of the same expansion. The augmented Eilenberger equations, defined on the common Fermi surface of the spin-independent dispersion, retain the band-off-diagonal interband coherence but neglect the helicity dependence of the Fermi-surface parameters. Conversely, we use the helicity-decoupled multiband Eilenberger equations only for the band-diagonal sector, thereby retaining the helicity-dependent Fermi momenta, velocities, and densities of states. Here, $\varepsilon_{\rm F}$ is the Fermi energy. For $\alpha\sim\Delta_0\ll\varepsilon_{\rm F}$, the relative Fermi-surface splitting is $O(\alpha/\varepsilon_{\rm F})=O(\delta)$, since the quasiclassical parameter is expressed as $\delta=\Delta_0/2\varepsilon_{\rm F}$. The effect of interband coherence is also of $O(\delta)$. Cross terms involving both effects are of $O(\delta^2)$, and can be neglected in the leading order of $\delta$. Therefore, the intraband and interband contributions ${\bf M}^{\rm intra}+{\bf M}^{\rm inter}$ can be evaluated separately with use of the complementary quasiclassical formulations in the order of $O(\delta)$ without double counting.

The leading-order multiband Eilenberger equations for each helicity band, $\lambda = \pm$, are given by
\begin{align}
\left[
i\varepsilon_n\hat{\tau}_3
-\hat{\Delta}_\lambda\hat{\tau}_3,
\hat{g}_\lambda
\right]
+i\hbar{\bf v}_{\rm F}^{(\lambda)}
\cdot{\bm\partial}\hat{g}_\lambda
=
\hat{0},
\label{eq-multiband_Eilenberger}
\end{align}
where
\begin{align}
\hat{g}_{\lambda}
&\equiv
\begin{bmatrix}
g_\lambda & -if_\lambda \\
-i\bar{f}_\lambda & -\bar{g}_\lambda
\end{bmatrix},
\qquad
\hat{\Delta}_{\lambda}
\equiv
\begin{bmatrix}
0 & \Delta_\lambda \\
-\bar{\Delta}_\lambda & 0
\end{bmatrix}.
\end{align}
Here, ${\bf v}_{\rm F}^{(\lambda)}$ is the Fermi velocity in helicity band $\lambda$. Solving Eq.~(\ref{eq-multiband_Eilenberger}) in the London limit, we obtain the quasiclassical Green's functions in the two helicity bands as
\begin{align}
g_\lambda
&=
\frac{\tilde{\varepsilon}_n^{(\lambda)}}
{\sqrt{\tilde{\varepsilon}_n^{(\lambda)2}+|\Delta_\lambda|^2}},
\qquad
f_\lambda
=
\frac{\Delta_\lambda}
{\sqrt{\tilde{\varepsilon}_n^{(\lambda)2}+|\Delta_\lambda|^2}},
\end{align}
where
\begin{subequations}
\begin{align}
\tilde{\varepsilon}_n^{(\lambda)}
&=
\varepsilon_n
+i{\bf v}_{\rm F}^{(\lambda)}\cdot{\bf p}_{\rm s},
\\
\Delta_+
&=
-\Delta_-{\rm e}^{2i\varphi_{\bf p}}
=
i\Delta{\rm e}^{i\varphi_{\bf p}}.
\end{align}
\end{subequations}

The intraband contribution to the magnetization is obtained by applying Eq.~(\ref{eq-M}) separately to each helicity band and summing the resulting contributions: 
\begin{align}
{\bf M}^{\rm intra}
=
i\pi k_{\rm B}T\mu_{\rm B}
\sum_{\lambda=\pm}N^{(\lambda)}(0)
\sum_{n=-\infty}^{\infty}
\left\langle
\frac{2}{\hbar}{\bf s}^{(\lambda)}g_\lambda
\right\rangle_{\rm F}^{(\lambda)}.
\end{align}
Here, $N^{(\lambda)}(0)$ and
$\langle\cdots\rangle_{\rm F}^{(\lambda)}$
denote the normal-state density of states at the Fermi level and the normalized Fermi-surface average in helicity band $\lambda$, respectively.
The spin expectation value ${\bf s}^{(\lambda)}$ is obtained from the diagonal component of the spin operator represented in the helicity basis:
\begin{align}
\underline{s}_i
\equiv
\underline{U}_{{\bf p}_{\rm F}}^\dagger
\frac{\hbar}{2}\underline{\sigma}_i
\underline{U}_{{\bf p}_{\rm F}}
\equiv
\begin{bmatrix}
s_i^{(+)} & s_i^{(+-)} \\
s_i^{(-+)} & s_i^{(-)}
\end{bmatrix}.
\end{align}

For cylindrical Fermi surfaces, the Fermi momenta, Fermi velocities, and normal-state densities of states in the two helicity bands are given by \cite{yip02}
\begin{subequations}
\begin{align}
p_{\rm F}^{(\lambda)}
&=
p_{\rm F}
\left[
\sqrt{
1+\left(\frac{\alpha}{2\varepsilon_{\rm F}}\right)^2}
-\lambda\frac{\alpha}{2\varepsilon_{\rm F}}
\right],
\\
v_{\rm F}^{(+)}
&=
v_{\rm F}^{(-)}
=
v_{\rm F}
\sqrt{
1+\left(\frac{\alpha}{2\varepsilon_{\rm F}}\right)^2},
\\
N^{(\lambda)}(0)
&=
\int_0^{2\pi}
\frac{p_{\rm F}^{(\lambda)}d\varphi_{\bf p}}
{(2\pi\hbar)^2v_{\rm F}^{(\lambda)}}
=
N(0)
\frac{p_{\rm F}^{(\lambda)}}
{mv_{\rm F}^{(\lambda)}},
\end{align}
\end{subequations}
where $p_{\rm F}$, $v_{\rm F}$, and $N(0)$ denote the Fermi momentum, Fermi velocity, and normal-state density of states, respectively, in the absence of Rashba spin-orbit coupling. For $\alpha\ll\varepsilon_{\rm F}$, the intraband contribution to the magnetization is then obtained as
\begin{align}
{\bf M}^{\rm intra}
=
-\frac{i\pi k_{\rm B}T\mu_{\rm B}N(0)}
{\varepsilon_{\rm F}}
\sum_{n=-\infty}^{\infty}
\left\langle
{\bf g}_{{\bf p}_{\rm F}}
\frac{\tilde{\varepsilon}_n}
{\sqrt{\tilde{\varepsilon}_n^2+|\Delta|^2}}
\right\rangle_{\rm F}.
\label{eq-M_intra-nonlinear}
\end{align}
Note that, for a cylindrical Fermi surface, the normalized Fermi-surface average is defined by
\begin{align}
\left\langle\cdots\right\rangle_{\rm F}
\equiv
\int_0^{2\pi}
\frac{d\varphi_{\bf p}}{2\pi}\cdots.
\end{align}
The total magnetization is obtained by summing intraband and interband contributions ${\bf M} = {\bf M}^{\rm intra}+{\bf M}^{\rm inter}$.

\subsection{Linear response limit}
Here, we consider the linear-response limit and obtain the current density and magnetization up to the first order in ${\bf p}_{\rm s}$ as
\begin{subequations}
\begin{align}
&{\bf j}
=
2\pi k_{\rm B}TeN(0) 
\sum_{n=-\infty}^{\infty}
\left\langle
{\bf v}_{\rm F}
\frac{|\Delta|^2}{(\varepsilon_n^2+|\Delta|^2)^{3/2}}
{\bf v}_{\rm F}\cdot{\bf p}_{\rm s}
\right\rangle_{\rm F}, 
\label{eq-j-linear} \\
&{\bf M}
=
2\pi k_{\rm B}T\mu_{\rm B}N(0)\sum_{n=-\infty}^{\infty} 
\Bigg\langle
\frac{{\bf g}_{{\bf p}_{\rm F}}}{2\varepsilon_{\rm F}}
\frac{|\Delta|^2}{(\varepsilon_n^2+|\Delta|^2)^{3/2}}
{\bf v}_{\rm F}\cdot{\bf p}_{\rm s} \notag \\
&-\frac{|\Delta|^2}
{(\varepsilon_n^2+\alpha^2+|\Delta|^2)
\sqrt{\varepsilon_n^2+|\Delta|^2}}
{\bf p}_{\rm s}\cdot\frac{\partial}{\partial{\bf p}_{\rm F}}{\bf g}_{{\bf p}_{\rm F}}
\Bigg\rangle_{\rm F}.
\label{eq-M-linear}
\end{align}
\end{subequations}
To this order, the superconducting gap can be replaced by its equilibrium value because its leading correction is of the second order in ${\bf p}_{\rm s}$. The equilibrium superconducting gap $|\Delta|$ is determined by
\begin{align}
\ln\frac{T}{T_{\rm c}}
+\pi k_{\rm B}T
\sum_{n=-\infty}^{\infty}
\left(
\frac{1}{|\varepsilon_n|}
-
\frac{1}
{\sqrt{\varepsilon_n^2+|\Delta|^2}}
\right)
=0.
\label{eq-gap-linear}
\end{align}

In the static limit, Amp\`ere's law in the presence of magnetization is written as 
\begin{align}
{\bm\nabla}\times
\left(
\frac{1}{\mu_0}{\bf B}-{\bf M}
\right)
={\bf j}, 
\label{eq-Maxwell}
\end{align}
where $\mu_0$ is the vacuum permeability. Since magnetization is of first order in the quasiclassical parameter $\delta$, it does not contribute to the zeroth-order Maxwell equation. Therefore, taking the curl of Eq.~(\ref{eq-Maxwell}) and using ${\bm\nabla}\cdot{\bf B}=0$, together with Eq.~(\ref{eq-j-linear}), we obtain the London equation \cite{lon61}
\begin{align}
{\bm\nabla}^2{\bf B}
=
\frac{1}{\lambda_{\rm L}^2}{\bf B},
\label{eq-London}
\end{align}
where $\lambda_{\rm L}$ is the London penetration depth at finite temperature, given by
\begin{align}
\lambda_{\rm L}
=
\lambda_0(1-Y)^{-1/2},
\end{align}
with $\lambda_0\equiv[\mu_0N(0)e^2v_{\rm F}^2]^{-1/2}$. Here, we note that two-dimensional isotropic Fermi surfaces have been assumed, and $Y$ is the Yosida function \cite{yos58},
\begin{align}
Y
=
1-\pi k_{\rm B}T
\sum_{n=-\infty}^{\infty}
\frac{|\Delta|^2}
{(\varepsilon_n^2+|\Delta|^2)^{3/2}}.
\end{align}
\begin{figure}[t]
\includegraphics[width=\linewidth]{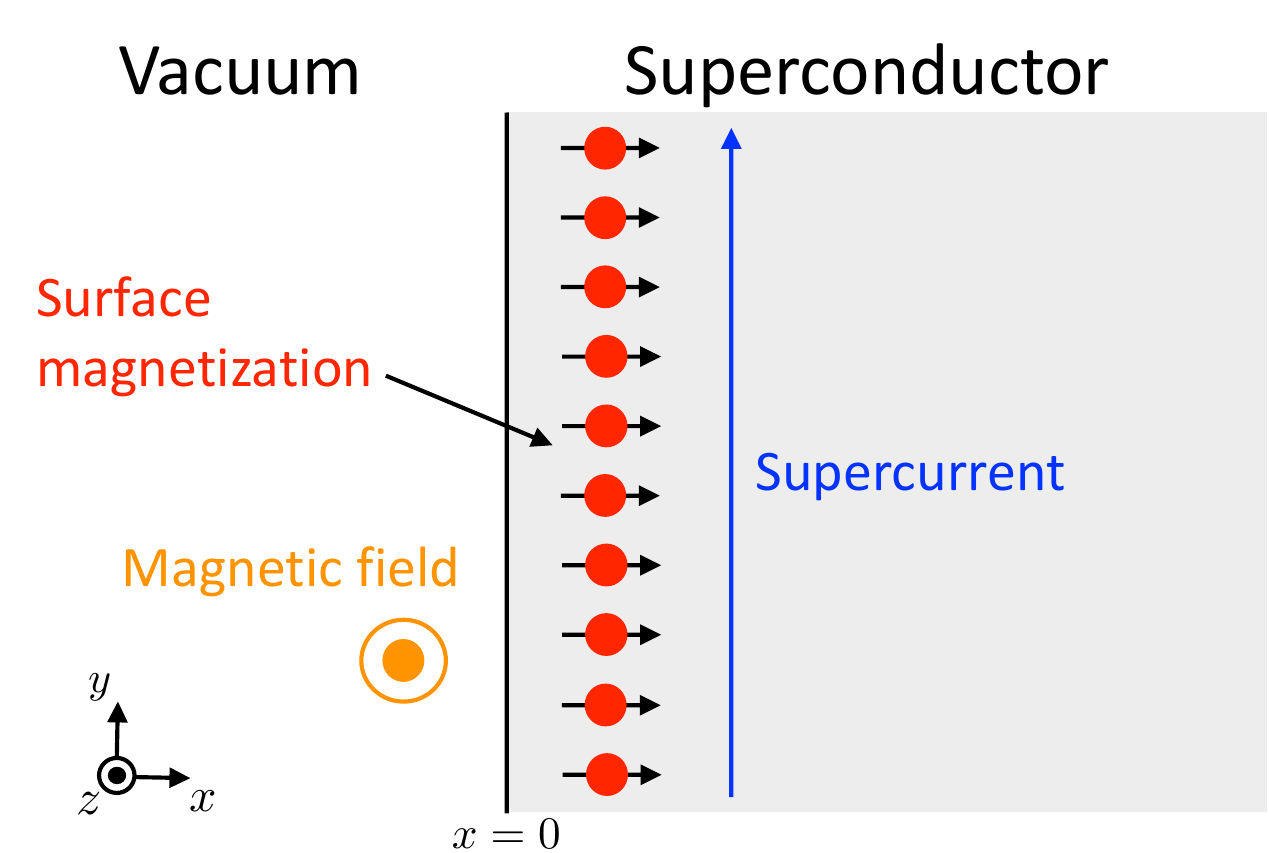}
\caption{
Schematic illustration of a semi-infinite Rashba superconductor subjected to a uniform magnetic field (orange arrow). Owing to the Rashba spin-orbit coupling, a spin magnetization polarized along the $x$ direction (red circles with arrows) is induced near the surface at $x=0$, where the Meissner current flows (blue arrow).
}
\label{semifinite_system}
\end{figure}
\begin{figure*}[t]
\includegraphics[width=\linewidth]{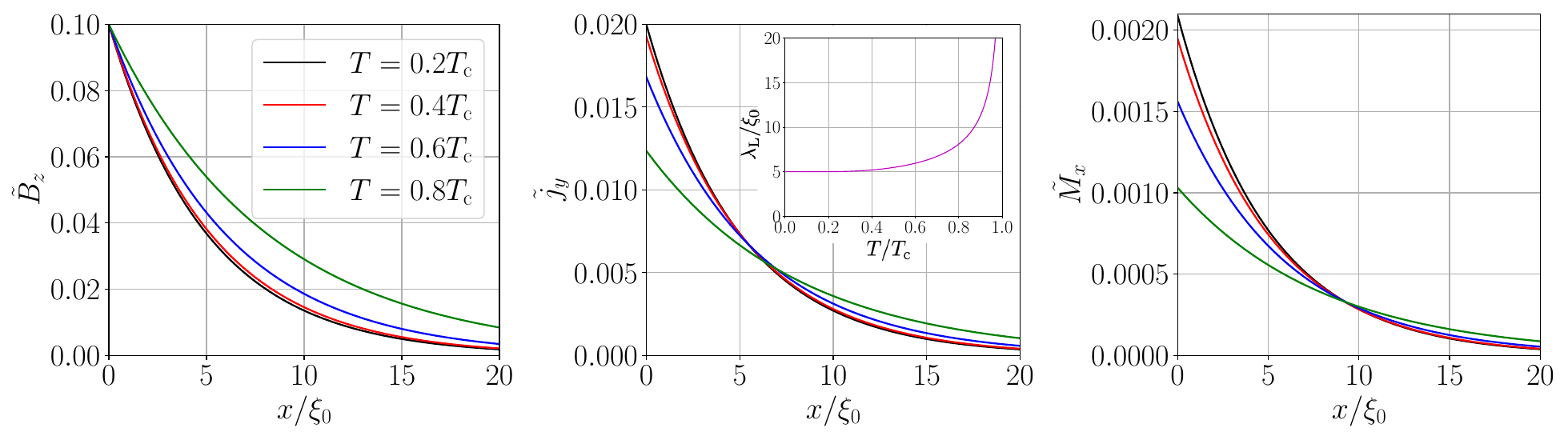}
\caption{
Spatial dependence of the magnetic field (left), current density (center), and magnetization (right) near the surface at $x=0$ in units of $\hbar/2|e|\xi_0^2$, $\hbar/2\mu_0|e|\xi_0^3$, and $\mu_{\rm B}N(0)\Delta_0$, respectively, for $\alpha=1.5\Delta_0$ at $T=0.2T_{\rm c}$, $0.4T_{\rm c}$, $0.6T_{\rm c}$, and $0.8T_{\rm c}$. The inset in the center panel shows the temperature dependence of the London penetration depth.
}
\label{spatial_dependence}
\end{figure*}
\begin{figure*}[t]
\includegraphics[width=0.65\linewidth]{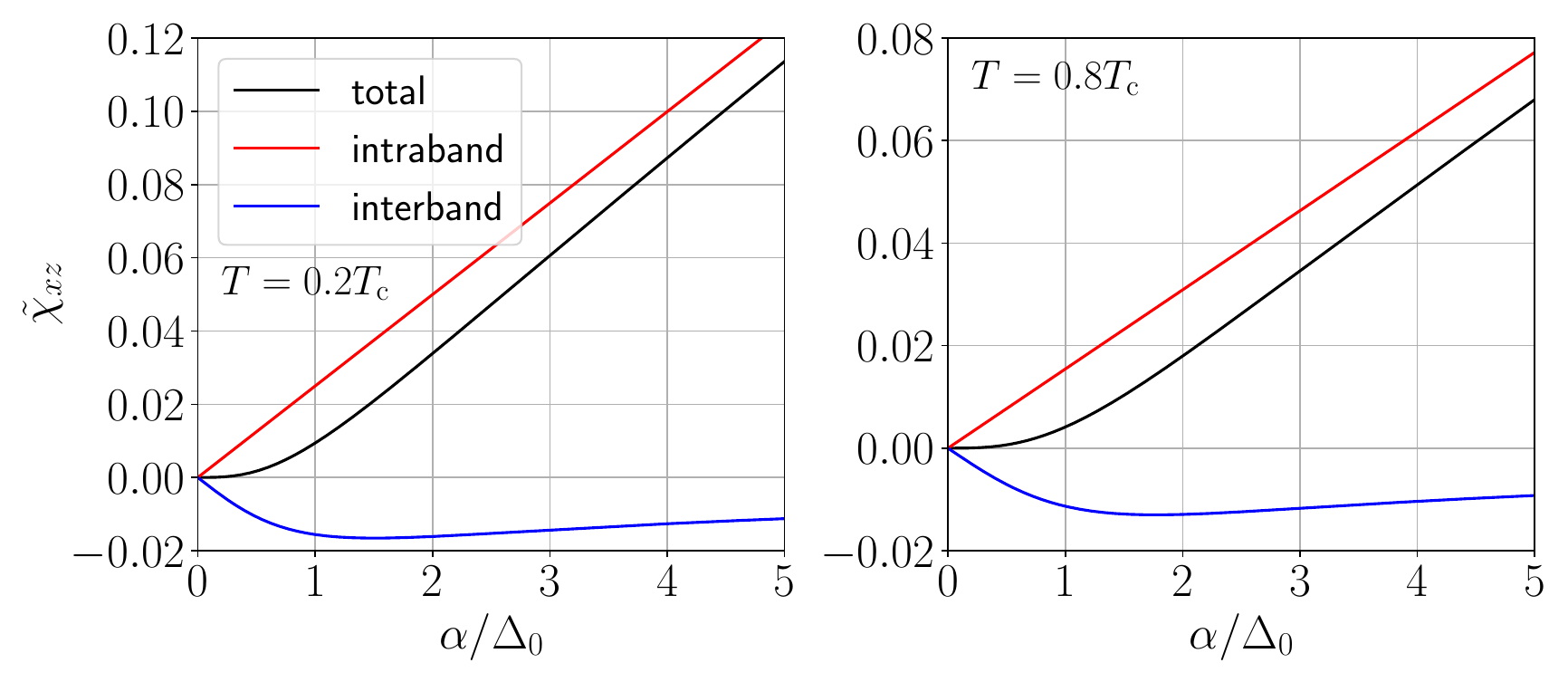}
\caption{
Intraband (red) and interband (blue) contributions to the normalized magnetic susceptibility $\tilde{\chi}_{xz}$, which originates from the Edelstein effect, together with their sum (black), as functions of $\alpha/\Delta_0$ at $T=0.2T_{\rm c}$ (left) and $0.8T_{\rm c}$ (right). The magnetic susceptibility is normalized by $2|e|\xi_0^2\mu_{\rm B}N(0)\Delta_0/\hbar$.
}
\label{magnetic_susceptibility}
\end{figure*}

\section{Numerical results\label{sec:v}}
\subsection{Linear Edelstein effect}
First, we focus on the linear response region and show the linear Edelstein effect. We consider an $s$-wave Rashba superconductor with cylindrical Fermi surfaces occupying the half space $x\geq0$, with a uniform magnetic field $B_0$ applied along the $z$ axis in the vacuum region $x<0$, as illustrated in Fig.~\ref{semifinite_system}. In this setup, the Meissner current that screens the applied magnetic field gives rise to magnetization through the Edelstein effect. Solving London equation Eq.~(\ref{eq-London}) for this geometry, we obtain the magnetic field ${\bf B}({\bf r})=B_z(x)\hat{\bf z}$ as
\begin{align}
B_z(x)=B_0{\rm e}^{-x/\lambda_{\rm L}}.
\label{eq-B-London}
\end{align}
Substituting Eq.~(\ref{eq-B-London}) into ${\bf j}=({\bm\nabla}\times{\bf B})/\mu_0$, we obtain the current density ${\bf j}({\bf r})=j_y(x)\hat{\bf y}$ as
\begin{align}
j_y(x)=\frac{B_0}{\mu_0\lambda_{\rm L}}{\rm e}^{-x/\lambda_{\rm L}}.
\label{eq-j-London}
\end{align}
Using this current distribution and ${\bf p}_{\rm s}=\mu_0e\lambda_{\rm L}^2 \,{\bf j}$, we finally obtain the magnetization ${\bf M}({\bf r})=M_x(x)\hat{\bf x}$ linear in ${\bf j}$ as 
\begin{align}
M_x(x)
&=
-\frac{\pi k_{\rm B}Te\mu_{\rm B}N(0)\lambda_{\rm L}}{p_{\rm F}}
B_0{\rm e}^{-x/\lambda_{\rm L}} \notag \\
&\times
\sum_{n=-\infty}^{\infty}
\frac{\alpha^3|\Delta|^2}
{(\varepsilon_n^2+\alpha^2+|\Delta|^2)
(\varepsilon_n^2+|\Delta|^2)^{3/2}}.
\label{eq-Mx-London}
\end{align}
The supercurrent along the {\it y} axis induces magnetization along the {\it x} axis, known as the superconducting Edelstein effect.

We choose the parameters $\kappa_0\equiv\lambda_0/\xi_0=5$, $\delta\equiv\hbar/p_{\rm F}\xi_0=0.01$, and $B_0=0.1(\hbar/2|e|\xi_0^2)$, and plot the following dimensionless quantities:
\begin{subequations}
\begin{align}
\tilde{B}_z(x)
&\equiv
\frac{B_z(x)}{\hbar/2|e|\xi_0^2}, \\
\tilde{j}_y(x)
&\equiv
\frac{j_y(x)}{\hbar/2\mu_0|e|\xi_0^3}, \\
\tilde{M}_x(x)
&\equiv
\frac{M_x(x)}{\mu_{\rm B}N(0)\Delta_0}, \\
\tilde{\chi}_{xz}
&\equiv
\frac{\chi_{xz}}{2|e|\xi_0^2\mu_{\rm B}N(0)\Delta_0/\hbar}
=
\frac{\tilde{M}_x(x)}{\tilde{B}_z(x)}.
\label{eq-chixz}
\end{align}
\end{subequations}
Here, $\chi_{xz}$ is the off-diagonal magnetic susceptibility defined by $\chi_{xz}\equiv M_x(x)/B_z(x)$, which characterizes the Edelstein effect arising from the Meissner current.

\begin{figure*}[t]
\includegraphics[width=\linewidth]{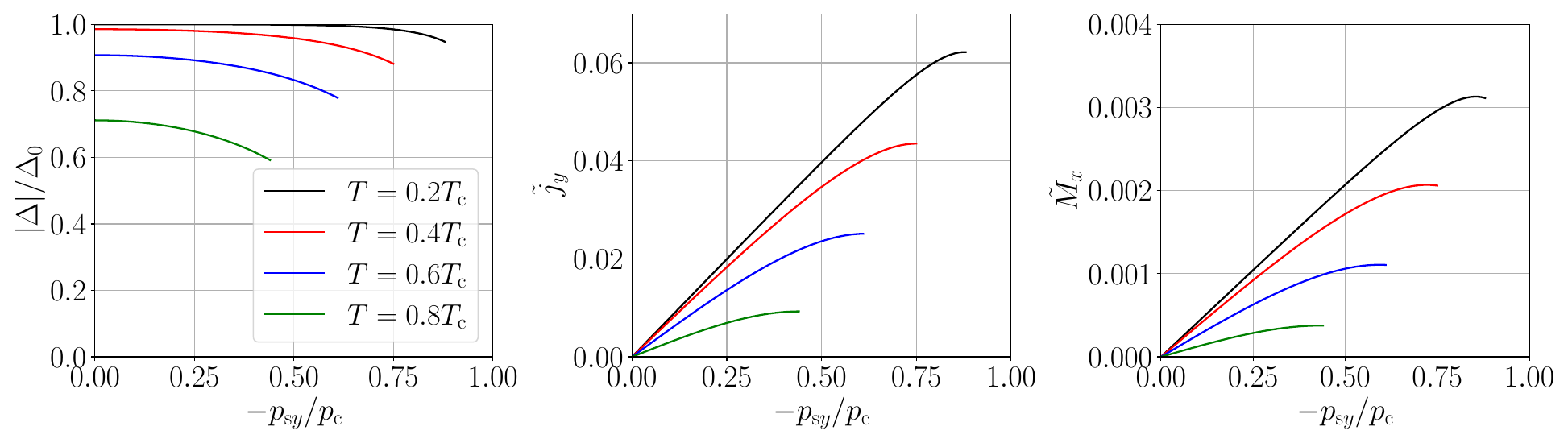}
\caption{
Superfluid-momentum dependence of the superconducting gap (left), current density (center), and magnetization (right), in units of $\Delta_0$, $\hbar/2\mu_0|e|\xi_0^3$, and $\mu_{\rm B}N(0)\Delta_0$, respectively, for $\alpha=1.5\Delta_0$ at $T=0.2T_{\rm c}$, $0.4T_{\rm c}$, $0.6T_{\rm c}$, and $0.8T_{\rm c}$. Here, the unit of superfluid momentum is $p_{\rm c}\equiv\Delta_0/v_{\rm F}$, and the superfluid momentum $p_{{\rm s}y}$ is treated as a local control parameter. In a semi-infinite geometry, however, it should be understood as $p_{{\rm s}y}=p_{{\rm s}y}(x;B_0)$, determined by the nonlinear Meissner screening profile, whose effective penetration depth depends on the surface magnetic field $B_0$ \cite{xu95,sau22}.
}
\label{ps-dependence}
\end{figure*}
\begin{figure*}[t]
\includegraphics[width=0.7\linewidth]{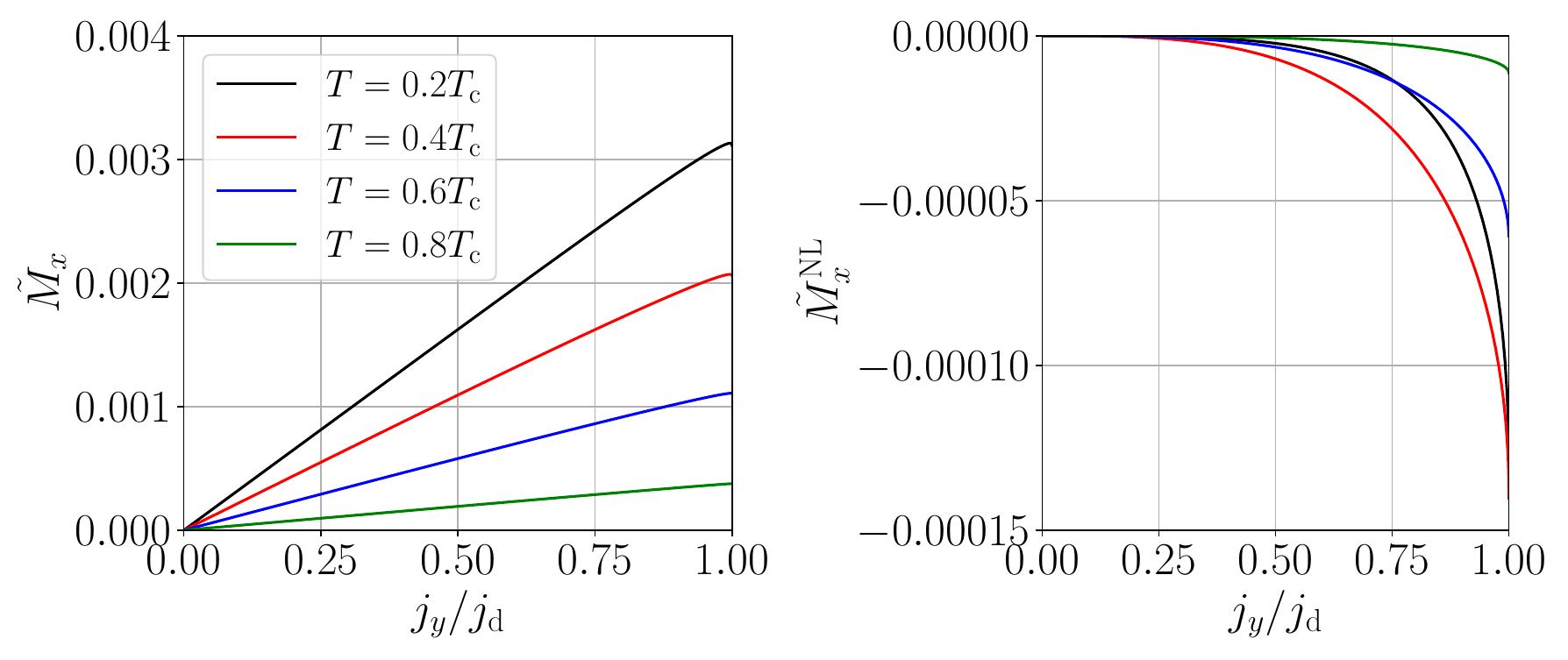}
\caption{
Current-density dependence of the magnetization normalized by $\mu_{\rm B}N(0)\Delta_0$ (left) and its nonlinear part (right) for $\alpha=1.5\Delta_0$ at $T=0.2T_{\rm c}$, $0.4T_{\rm c}$, $0.6T_{\rm c}$, and $0.8T_{\rm c}$. The nonlinear part is defined as $\tilde{M}_x^{\rm NL}\equiv\tilde{M}_x-\tilde{M}_x^{\rm L}$, where $\tilde{M}_x^{\rm L}$ denotes the part of $\tilde{M}_x$ linear in the current density $j_y$. The unit of $j_y$ is the depairing current density $j_{\rm d}$.
}
\label{supercurrent_dependence}
\end{figure*}

Figure~\ref{spatial_dependence} shows the spatial dependence of the magnetic field, current density, and magnetization near the surface at $x=0$ for $\alpha=1.5\Delta_0$ at temperatures $T=0.2T_{\rm c}$, $0.4T_{\rm c}$, $0.6T_{\rm c}$, and $0.8T_{\rm c}$. All these quantities decay exponentially into the bulk on the scale of the London penetration depth $\lambda_{\rm L}$. As the temperature increases, $\lambda_{\rm L}$ increases, so that the magnetic field, current density, and magnetization extend more deeply into the superconductor. The magnetic field and current density reproduce the conventional London screening behavior.

Figure~\ref{magnetic_susceptibility} shows the intraband and interband contributions to the off-diagonal magnetic susceptibility, together with their sum, as functions of the spin-orbit coupling strength $\alpha$ at $T=0.2T_{\rm c}$ and $0.8T_{\rm c}$. The intraband contribution is proportional to $\alpha$, whereas the interband contribution exhibits a maximum at an intermediate spin-orbit coupling strength. For small $\alpha$, the interband contribution increases approximately linearly with $\alpha$, while for large $\alpha$, it decreases as $1/\alpha$ owing to the increasing energy splitting between the helicity bands. The maximum appears around $\alpha\simeq1.5\Delta_0$. For each Matsubara component, the characteristic scale of spin-orbit coupling is set not simply by $|\Delta|$, but by $\sqrt{\varepsilon_n^2+|\Delta|^2}$. The summation over finite Matsubara frequencies therefore shifts and broadens the maximum relative to the naive estimate $\alpha\sim|\Delta|$.

A closely related decomposition of the linear superconducting Edelstein response was obtained in the quantum geometric pair potential (QGPP) formulation of Ref.~\cite{dai24}. For the spin-singlet two-helicity-band model considered here, conventional band-dispersion contribution and quantum-geometric QGPP contribution in their formulation correspond to our intraband and interband terms, respectively. In their generalized band representation, the superflow-induced interband coherence is encoded in an off-diagonal QGPP and the associated mixed quantum metric. In the unperturbed-helicity representation used in this paper, the same interband physics appears through the anomalous-velocity term involving the momentum-space Berry connection.

For the present isotropic parabolic Rashba model with a cylindrical Fermi surface, the zero-temperature Matsubara integration of Eq.~(\ref{eq-M-linear}) gives
\begin{align}
\left.
\frac{M_x^{\rm inter}}{M_x^{\rm intra}}
\right|_{T=0}
&=
-\frac{
\displaystyle
\int_{-\infty}^{\infty}d\varepsilon
\frac{\Delta_0^2}
{\sqrt{\varepsilon^2+\Delta_0^2}
(\varepsilon^2+\Delta_0^2+\alpha^2)}
}{
\displaystyle
\int_{-\infty}^{\infty}d\varepsilon
\frac{\Delta_0^2}
{(\varepsilon^2+\Delta_0^2)^{3/2}}
}
\notag\\
&=
-I(\Delta_0/\alpha),
\end{align}
where
\begin{align}
I(x)
\equiv
\frac{x^2}{\sqrt{1+x^2}}
\tanh^{-1}
\left(
\frac{1}{\sqrt{1+x^2}}
\right),
\end{align}
in agreement with the zero-temperature ratio between the quantum-geometric and conventional contributions, $\chi_{\rm g}^{h_xq_y}/\chi_{\rm dia}^{h_xq_y}=-I(\Delta_0/\alpha)$, obtained in Ref.~\cite{dai24}. For $\alpha\ll\Delta_0$, we have
\begin{align}
I(\Delta_0/\alpha)
=
1-\frac{2}{3}
\left(\frac{\alpha}{\Delta_0}\right)^2
+O\left[
\left(\frac{\alpha}{\Delta_0}\right)^4
\right].
\end{align}
Since $M_x^{\rm intra}\propto\alpha$, the total magnetization $M_x^{\rm intra}+M_x^{\rm inter}=M_x^{\rm intra}[1-I(\Delta_0/\alpha)]$ therefore begins with a term proportional to $\alpha^3$.

The cubic scaling of the supercurrent-induced magnetization in the limit of weak spin-orbit coupling can also be understood within the SU(2) gauge-field formulation~\cite{kon15}. For spatially uniform Rashba spin-orbit coupling, ${\cal A}_i\propto\alpha$, giving
\begin{align}
{\cal F}_{ij}
&=-i[{\cal A}_i,{\cal A}_j]\propto\alpha^2,
\\
{\cal J}_j
&=\tilde{\nabla}_i{\cal F}_{ij}
=-i[{\cal A}_i,{\cal F}_{ij}]\propto\alpha^3,
\end{align}
where ${\cal J}_j={\cal J}_j^a\sigma^a/2$ is proportional to the equilibrium spin current, with $\sigma^a$ ($a=x,y,z$) denoting the Pauli matrices in their notation. Near $T_c$, the leading magnetoelectric contribution to the free energy has the form $F_{\rm ME}\propto{\cal A}_0^a{\cal J}_i^a p_{{\rm s},i}$ [Eq.~(3.6) of Ref.~\cite{kon15}]. Introducing ${\cal A}_0^a$ as an auxiliary Zeeman source, differentiating the free energy with respect to it at fixed superfluid momentum, and then setting all components ${\cal A}_0^a$ to zero gives $M_a\propto{\cal J}_i^a p_{{\rm s},i}$. Thus, in the limit of weak spin-orbit coupling, the magnetization linear in the superfluid momentum scales as $\alpha^3$. This scaling is consistent with the cancellation of the $\alpha$-linear intraband and interband contributions in our calculation, as shown in Fig.~\ref{magnetic_susceptibility}.

\subsection{Nonlinear Edelstein effect}
Using the formulation, Eqs.~(\ref{eq-j-nonlinear}), (\ref{eq-M_inter-nonlinear}), (\ref{eq-gap-nonlinear}), and (\ref{eq-M_intra-nonlinear}), we calculate the current density, superconducting gap, and magnetization as functions of the superfluid momentum ${\bf p}_{\rm s}=p_{{\rm s}y}\hat{\bf y}$. From these results, we determine the depairing current density $j_{\rm d}$ and obtain the current dependence of magnetization, ${\bf M}=M_x(j_y)\hat{\bf x}$, an expression often referred to as the Edelstein effect. We use the same parameters $\kappa_0\equiv\lambda_0/\xi_0=5$ and $\delta\equiv\hbar/p_{\rm F}\xi_0=0.01$ as in the previous subsection. In contrast to the previous subsection, nonlinear effects with respect to the superfluid momentum is taken into account.

Figure~\ref{ps-dependence} shows the superconducting gap, current density, and magnetization as functions of the superfluid momentum. Following Ref.~\cite{kub22}, we define the depairing current density $j_{\rm d}$ as the maximum of $j_y(p_{{\rm s}y})$. We obtain $j_{\rm d}/(\hbar/2\mu_0|e|\xi_0^3)=6.22\times10^{-2}$, $4.35\times10^{-2}$, $2.51\times10^{-2}$, and $9.27\times10^{-3}$ at $T=0.2T_{\rm c}$, $0.4T_{\rm c}$, $0.6T_{\rm c}$, and $0.8T_{\rm c}$, respectively. A nonequilibrium calculation further shows that, in a dirty superconductor near $T_{\rm c}$, the order parameter eventually vanishes when a current density exceeding $j_{\rm d}$ is applied \cite{she20}. In the present static theory, both superconducting gap $\Delta$ and current density $j_y$ exhibit pronounced nonlinear dependences on $p_{{\rm s}y}$ near the depairing current $j_{\rm d}$, in agreement with the static current-carrying solutions of Ref.~\cite{kub22}. Our calculation demonstrates that magnetization $M_x$ shows a qualitatively similar dependence on $p_{{\rm s}y}$.

From the results in Fig.~\ref{ps-dependence}, we obtain the current-density dependence of magnetization by eliminating $p_{{\rm s}y}$ between $j_y(p_{{\rm s}y})$ and $M_x(p_{{\rm s}y})$ on the branch where $j_y$ increases monotonically with $p_{{\rm s}y}$ up to the depairing current density $j_{\rm d}$.  In Fig.~\ref{supercurrent_dependence}, we find that $M_x$ remains nearly linear in $j_y$ over a wide current range, whereas the nonlinear contribution $M_x^{\rm NL}\equiv M_x-M_x^{\rm L}$ becomes appreciable near the depairing current density $j_{\rm d}$. Here, the linear component is $M_x^{\rm L}\equiv\partial M_x/\partial j_y|_{j_y=0}j_y$, where the slope is evaluated from the linear-response expression.

Let us discuss the origin of the nonlinear Edelstein effect demonstrated in Fig.~\ref{supercurrent_dependence}. The $p_{{\rm s}y}$ dependence of the intraband magnetization $M_x^{\rm intra}$ is identical to that of $j_y$. Indeed, Eqs.~(\ref{eq-j-nonlinear}) and (\ref{eq-M_intra-nonlinear}) give the linear relationship 
\begin{align}
\tilde{M}_x^{\rm intra}
=
\frac{\delta}{2}\frac{\alpha}{\Delta_0}\tilde{j}_y.
\end{align}
Therefore, $M_x^{\rm intra}$ is exactly linear in $j_y$ within the present theory. By contrast, the $p_{{\rm s}y}$ dependence of the interband magnetization $M_x^{\rm inter}$ differs from that of $j_y$, because $j_y$ and $M_x^{\rm inter}$ are determined by the diagonal and off-diagonal components of the Green's functions, respectively. It follows that the nonlinear magnetization $M_x^{\rm NL}$ originates entirely from the interband component $M_x^{\rm inter}$.

The right panel of Fig.~\ref{supercurrent_dependence} shows that, at a low temperature $T=0.2T_{\rm c}$, the nonlinear correction remains small over most of the current range and increases rapidly in the vicinity of the depairing current. At an intermediate temperature $T=0.4T_{\rm c}$, it becomes appreciable at lower currents and remains relatively large over a broader current range. This behavior cannot be attributed solely to the superflow-induced suppression of the superconducting gap. The gap suppression almost equivalently influences both $M_x^{\rm inter}$ and $j_y$, and its effect is therefore largely removed when considering the deviation of $M_x$ from its linear dependence on $j_y$. The nonlinear correction instead reflects the difference between the $p_{{\rm s}y}$ dependences of the interband magnetization $M_x^{\rm inter}$ and the supercurrent density $j_y$. At low temperatures, these dependences remain nearly proportional until strong pair breaking occurs close to the depairing current. In contrast, at intermediate temperatures, the proportionality begins to deviate at smaller currents, producing the broader enhancement of nonlinearlity observed at $T=0.4T_{\rm c}$. At higher temperatures, the superconducting gap is more strongly suppressed by the superflow, and the magnitude of the interband magnetization is reduced, resulting in a smaller nonlinear correction.

\section{Summary\label{sec:vi}}
We derived augmented Eilenberger equations for noncentrosymmetric superconductors with general antisymmetric spin-orbit coupling and applied them to an $s$-wave Rashba superconductor. By solving  the equations in the London and linear-response limits, including the anomalous-velocity terms arising from the momentum derivative of the Rashba spin-orbit coupling, we showed a magnetization perpendicular to both the Meissner screening current and the magnetic field, which is regarded as the surface Edelstein effect due to the Meissner current. The interband contribution to the surface magnetization is enhanced at low temperatures and exhibits a nonmonotonic dependence on the Rashba spin-orbit coupling, reaching a maximum when the Rashba spin splitting is comparable to the superconducting gap. Our formulation also makes explicit the role of the Berry connection associated with the momentum dependence of the helicity basis, demonstrating that the interband Edelstein response in spin-singlet Rashba superconductors originates from interband matrix elements between the Rashba helicity bands. We further calculated the intraband contribution using the multiband Eilenberger equations and found that it dominates the linear-response magnetization.

We also investigated the supercurrent-induced magnetization beyond the linear-response regime, namely, the nonlinear Edelstein effect. Within the clean $s$-wave Rashba model considered in this paper, the intraband magnetization has the same superfluid-momentum dependence as the supercurrent and is therefore strictly linear in the supercurrent. In contrast, the interband contribution exhibits a distinct supercurrent dependence and produces a nonlinear magnetization that becomes appreciable near the depairing current density. Consequently, our results highlight the nonlinear Edelstein effect as a sensitive probe of the interband contribution, regarded as a quantum geometric term, in the clean isotropic Rashba superconductors.

Although this paper focused on clean $s$-wave superconductors, impurity effects and $d$-wave pairing states can be treated using the methods developed in Refs.~\cite{yip92,xu95,sau22}. Realistic Fermi surfaces and material-dependent parameters, such as those considered in Ref.~\cite{hig23}, can also be incorporated to estimate the magnitude of the magnetization in candidate materials. In addition, the Lorentz force acting on the supercurrent, which appears in Eq.~(\ref{eq-augmented_Eilenberger}), is known to induce the equilibrium Hall effect \cite{kit09}. The augmented Eilenberger framework therefore describes both the surface magnetization associated with the Edelstein effect and the surface charge associated with the equilibrium Hall effect in a unified framework. The latter has been observed experimentally \cite{bok68,mor71}.

\begin{acknowledgments}
The authors thank Yoichi Higashi, Taiki Matsushita, Yuma Hirobe, J. A. Sauls, and Wei-Ting Lin for helpful discussions. This work is supported by JSPS KAKENHI (Grant Numbers JP22H04933, JP23K17353, JP24K21530, JP24H00007, JP25H01249, JP26H02016).
\end{acknowledgments}

\appendix
\section{Charge density and magnetization\label{AppA}}
The electron density $n$ is given by \cite{kit10}
\begin{align}
n=\hbar\int\frac{d^3p\,d\varepsilon}{(2\pi\hbar)^4}
{\rm Tr}\,\underline{A}\underline{\phi},
\end{align}
where $\underline{A}(\varepsilon,{\bf p},{\bf r},t)$ is the spectral function and $\underline{\phi}(\varepsilon,{\bf p},{\bf r},t)$ is the distribution function in spin space. The relations between the Keldysh Green's functions and the spectral and distribution functions in spin space are given by
\begin{subequations}
\begin{align}
&\underline{G}^{\rm R}(\varepsilon,{\bf p},{\bf r},t)
-\underline{G}^{\rm A}(\varepsilon,{\bf p},{\bf r},t) 
=-i\underline{A}(\varepsilon,{\bf p},{\bf r},t), \\
&\underline{G}^{\rm K}(\varepsilon,{\bf p},{\bf r},t)
=-i\underline{A}(\varepsilon,{\bf p},{\bf r},t)
\left[
\underline{\sigma}_0-2\underline{\phi}(\varepsilon,{\bf p},{\bf r},t)
\right],
\end{align}
\end{subequations}
where $\underline{G}^{\rm R}$, $\underline{G}^{\rm A}$, and $\underline{G}^{\rm K}$ are the retarded, advanced, and Keldysh components, respectively. Using these relations, the electron density can be rewritten as
\begin{align}
n=-\frac{i\hbar}{2}\int\frac{d^3p\,d\varepsilon}{(2\pi\hbar)^4}
{\rm Tr}\left(
\underline G^{\rm K}-\underline G^{\rm R}+\underline G^{\rm A}
\right).
\end{align}
The charge density $\rho=e(n-n_{\rm eq})$ is then given by
\begin{align}
\rho
=&-\frac{i\hbar e}{2}\int\frac{d^3p\,d\varepsilon}{(2\pi\hbar)^4}
{\rm Tr}\bigl(
\underline G^{\rm K}-\underline G^{\rm R}+\underline G^{\rm A} \notag \\
& \hspace{20mm} \ \ \ 
-\underline G_{\rm eq}^{\rm K}
+\underline G_{\rm eq}^{\rm R}
-\underline G_{\rm eq}^{\rm A}
\bigr),
\end{align}
where quantities with the subscript ``eq'' are evaluated in equilibrium. We use the sum rule for the spectral function \cite{kit10},
\begin{align}
\int\frac{d\varepsilon}{2\pi}
\underline A
=
\underline \sigma_0, 
\label{eq-sum_rule_A}
\end{align}
which gives
\begin{align}
\int\frac{d\varepsilon}{2\pi}
i\left(\underline G^{\rm R}-\underline G^{\rm A}\right)
=
\int\frac{d\varepsilon}{2\pi}
i\left(\underline G_{\rm eq}^{\rm R}-\underline G_{\rm eq}^{\rm A}\right).
\end{align}
Since the single-particle energy is dominant and the external-field terms can be neglected in the high-energy Green's functions, we may approximate $\underline G^{\rm K}\approx\underline G_{\rm eq}^{\rm K}$ in the high-energy region. Thus, $\rho$ can be written as
\begin{align}
\rho=-\frac{i\hbar e}{2}
\int_{-\varepsilon_{\rm c}}^{\varepsilon_{\rm c}}
\frac{d\varepsilon}{2\pi\hbar}
\int\frac{d^3p}{(2\pi\hbar)^3}
{\rm Tr}\left(
\underline G^{\rm K}-\underline G_{\rm eq}^{\rm K}
\right), 
\label{eq-rho}
\end{align}
where $\varepsilon_{\rm c}$ is the energy cutoff. The momentum integral in Eq.~(\ref{eq-rho}) is rewritten as
\begin{align}
\int\frac{d^3p}{(2\pi\hbar)^3}\cdots
=
\int_{-\infty}^\infty d\xi_{\bf p}
N(\xi_{\bf p})
\langle\cdots\rangle_{\xi_{\bf p}},
\end{align}
where $\langle\cdots\rangle_{\xi_{\bf p}}$ denotes the average over the constant-$\xi_{\bf p}$ surface, normalized as $\langle1\rangle_{\xi_{\bf p}}=1$. We use $N(\xi_{\bf p})\approx N(0)$ and $\langle\cdots\rangle_{\xi_{\bf p}}\approx\langle\cdots\rangle_{\rm F}$, and introduce the quasiclassical Green's functions as
\begin{align}
&
\begin{bmatrix}
\underline{g}^{\rm R}(\varepsilon,{\bf p}_{\rm F},{\bf r},t)  
&\underline{g}^{\rm K}(\varepsilon,{\bf p}_{\rm F},{\bf r},t)   \\
\underline{0}
&\underline{g}^{\rm A}(\varepsilon,{\bf p}_{\rm F},{\bf r},t) 
\end{bmatrix} \notag \\
&\equiv
{\rm P} \int_{-\infty}^\infty \frac{d \xi_{\bf p}}{\pi}
i 
\begin{bmatrix}
\underline{G}^{\rm R}(\varepsilon,{\bf p},{\bf r},t)  
&\underline{G}^{\rm K}(\varepsilon,{\bf p},{\bf r},t)   \\
\underline{0}
&\underline{G}^{\rm A}(\varepsilon,{\bf p},{\bf r},t) 
\end{bmatrix}. 
\label{eq-gK}
\end{align}
Then, the charge density is given by
\begin{align}
\rho=-\frac{eN(0)}{4}
\int_{-\varepsilon_{\rm c}}^{\varepsilon_{\rm c}}
d\varepsilon
\left\langle
{\rm Tr}\left(
\underline g^{\rm K}-\underline g_{\rm eq}^{\rm K}
\right)
\right\rangle_{\rm F}.
\end{align}
Since the equilibrium Keldysh Green's function is given by $\underline G_{\rm eq}^{\rm K}=-i\underline A(1-2\underline \phi_{\rm eq})$ with the Fermi-Dirac distribution function $\underline \phi_{\rm eq}=({\rm e}^{\varepsilon/k_{\rm B}T}+1)^{-1}\underline\sigma_0$ \cite{kit10}, ${\rm Tr}\underline{g}_{\rm eq}^{\rm K}$ is calculated as
\begin{align}
{\rm Tr}\underline{g}_{\rm eq}^{\rm K}
=
2{\rm Re}{\rm Tr}\underline g_{\rm eq}^{\rm R}
\tanh\frac{\varepsilon}{2k_{\rm B}T},
\end{align}
where we have used $\underline{g}^{\rm A}=-\underline{g}^{\rm R}{}^\dagger$ \cite{kit01,kit09}. Since the density of states, $\langle{\rm Re}{\rm Tr}\underline g_{\rm eq}^{\rm R}\rangle_{\rm F}$, is an even function of $\varepsilon$, while $\tanh(\varepsilon/2k_{\rm B}T)$ is odd, the equilibrium contribution vanishes upon integration over the symmetric energy interval. We therefore obtain the charge density as
\begin{align}
\rho=-\frac{eN(0)}{4}
\int_{-\varepsilon_{\rm c}}^{\varepsilon_{\rm c}}d\varepsilon 
\left\langle
{\rm Tr}\,\underline g^{\rm K}
\right\rangle_{\rm F}. 
\label{eq-rho2}
\end{align}

Following the same procedure as in the derivation of Eq.~(\ref{eq-rho2}), we calculate the magnetization. The magnetization ${\bf M}$ is given by
\begin{align}
{\bf M}=
-\frac{2\mu_{\rm B}}{\hbar}
\hbar\int\frac{d^3p\,d\varepsilon}{(2\pi\hbar)^4}
{\rm Tr}\,
\frac{\hbar}{2}
\underline{\bm\sigma}\,
\underline{A}\underline{\phi}.
\end{align}
Using the Keldysh Green's functions, it can be expressed as
\begin{align}
{\bf M}=
-\frac{2\mu_{\rm B}}{\hbar}
\left[
-\frac{i\hbar}{2}
\int\frac{d^3p\,d\varepsilon}{(2\pi\hbar)^4}
{\rm Tr}\,
\frac{\hbar}{2}
\underline{\bm\sigma}
\left(
\underline G^{\rm K}
-\underline G^{\rm R}
+\underline G^{\rm A}
\right)
\right].
\end{align}
Using Eqs.~(\ref{eq-sum_rule_A}) and (\ref{eq-gK}), we obtain the expression for the magnetization as
\begin{align}
{\bf M}=
\frac{\mu_{\rm B}N(0)}{4}
\int_{-\varepsilon_{\rm c}}^{\varepsilon_{\rm c}}d\varepsilon 
\left\langle
{\rm Tr}\,
\underline {\bm\sigma}\underline g^{\rm K}
\right\rangle_{\rm F}. 
\label{eq-M-Keldysh}
\end{align}
We obtain the equilibrium magnetization, Eq.~(\ref{eq-M}), from Eq.~(\ref{eq-M-Keldysh}) by analytic continuation to the Matsubara formalism, using $\underline{g}^{\rm K}=(\underline{g}^{\rm R}-\underline{g}^{\rm A})\tanh(\varepsilon/2k_{\rm B}T)$ and the residue theorem.

\section{Equation of continuity\label{AppB}}
The main part of the quasiclassical equations of superconductivity for the Keldysh component corresponding to Eq.~(\ref{eq-augmented_Eilenberger_anomalous_velocity}) is given by \cite{kit01}
\begin{align}
&\left[ 
\varepsilon\hat{\tau}_3
-{\bf g}_{{\bf p}_{\rm F}}\cdot\hat{\bm\sigma}\hat{\tau}_3
-\hat{\Delta}\hat{\tau}_3, 
\hat{g}^{\rm K} 
\right]_\circ
+\frac{i\hbar}{2}
\left\{
\hat{\bf v}_{\rm F}, 
{\bm\partial}\hat{g}^{\rm K} 
\right\} \notag \\
& \ \ \ 
+\frac{i\hbar}{2}e{\bf v}_{\rm F}\cdot{\bf E}
\frac{\partial}{\partial\varepsilon}
\left\{
\hat{\tau}_3,\hat{g}^{\rm K} 
\right\}
=\hat{0},  
\end{align}
where
\begin{align}
\hat{\bf v}_{\rm F}
\equiv
{\bf v}_{\rm F}\hat{1}
+\frac{\partial}{\partial{\bf p}_{\rm F}}
\left(
{\bf g}_{{\bf p}_{\rm F}}\cdot\hat{\bm\sigma}\hat{\tau}_3
\right).
\end{align}
For the gauge ${\bf E}=-\partial{\bf A}/\partial t$ and $\Phi=0$, this equation can be rewritten as \cite{esc99}
\begin{align}
&\left[ 
\left(
\varepsilon
+e{\bf v}_{\rm F}\cdot{\bf A}
\right)\hat{\tau}_3
-{\bf g}_{{\bf p}_{\rm F}}\cdot\hat{\bm\sigma}\hat{\tau}_3
-\hat{\Delta}\hat{\tau}_3, 
\hat{g}^{\rm K} 
\right]_\circ \notag \\
& \ \ \ 
+\frac{i\hbar}{2}
\left\{
\hat{\bf v}_{\rm F}, 
{\bm\nabla}\hat{g}^{\rm K} 
\right\}
=\hat{0}. 
\label{eq-augmented_Eilenberger_anomalous_velocity-Keldysh}
\end{align}
Here, ${\bf E}$ is the electric field, $\Phi$ is the scalar potential, $[\hat{a},\hat{b}]_\circ$ is defined by $[\hat{a},\hat{b}]_\circ=\hat{a}\circ\hat{b}-\hat{b}\circ\hat{a}$, and the operator $\circ$ is defined by
\begin{align}
&\hat{a}(\varepsilon,t)\circ \hat{b}(\varepsilon,t) \notag \\
&\equiv 
{\rm e}^{
\frac{i\hbar}{2} 
\left( 
\frac{\partial}{\partial\varepsilon}\frac{\partial}{\partial t'}
-\frac{\partial}{\partial t}\frac{\partial}{\partial\varepsilon'} 
\right)
}
\hat{a}(\varepsilon,t)\hat{b}(\varepsilon',t')
\bigg|_{\varepsilon'=\varepsilon,\ t'=t}. 
\end{align}
The matrix $\hat{g}^{\rm K}$ is expressed as
\begin{align}
&\hat{g}^{\rm K}(\varepsilon,{\bf p}_{\rm F},{\bf r},t) 
\equiv {\rm P} \int_{-\infty}^\infty \frac{d \xi_{\bf p}}{\pi} 
i\hat{\tau}_3\hat{G}^{\rm K}(\varepsilon,{\bf p},{\bf r},t) \notag \\
&\equiv
\begin{bmatrix}
\underline{g}^{\rm K}(\varepsilon,{\bf p}_{\rm F},{\bf r},t) 
&-i\underline{f}^{\rm K}(\varepsilon,{\bf p}_{\rm F},{\bf r},t) \\
i\underline{f}^{\rm K}{}^*(-\varepsilon,-{\bf p}_{\rm F},{\bf r},t) 
&\underline{g}^{\rm K}{}^*(-\varepsilon,-{\bf p}_{\rm F},{\bf r},t)
\end{bmatrix}. 
\end{align}
It is convenient to introduce the barred functions in the Keldysh formalism as $\bar{X}(\varepsilon,{\bf p}_{\rm F},{\bf r},t) \equiv X^*(-\varepsilon,-{\bf p}_{\rm F},{\bf r},t)$.

The $(1,1)$ component of Eq.~(\ref{eq-augmented_Eilenberger_anomalous_velocity-Keldysh}) is given by
\begin{align}
&i\hbar\frac{\partial\underline{g}^{\rm K}}{\partial t} 
+i\hbar{\bf v}_{\rm F}\cdot{\bm\nabla}\underline{g}^{\rm K}
+i\hbar e{\bf v}_{\rm F}\cdot{\bf E}
\frac{\partial\underline{g}^{\rm K}}{\partial\varepsilon} \notag \\
& \ \ \ 
-\left({\bf g}_{{\bf p}_{\rm F}}\cdot\underline{\bm\sigma}\right)
\underline{g}^{\rm K}
+\underline{g}^{\rm K}
\left({\bf g}_{{\bf p}_{\rm F}}\cdot\underline{\bm\sigma}\right)
+i\underline{\Delta}\,\underline{\bar{f}}^{\rm K}
+i\underline{f}^{\rm K}\,\underline{\bar{\Delta}} \notag \\
& \ \ \ 
+\frac{i\hbar}{2}
\frac{\partial}{\partial{\bf p}_{\rm F}}
\left(
{\bf g}_{{\bf p}_{\rm F}}\cdot\underline{\bm\sigma}
\right)
\cdot
{\bm\nabla}\underline{g}^{\rm K}
+\frac{i\hbar}{2}
{\bm\nabla}\underline{g}^{\rm K}
\cdot
\frac{\partial}{\partial{\bf p}_{\rm F}}
\left(
{\bf g}_{{\bf p}_{\rm F}}\cdot\underline{\bm\sigma}
\right) \notag \\
& \ \ \ 
+\frac{\hbar}{2}
\frac{\partial\underline{\Delta}}{\partial t}
\frac{\partial\underline{\bar{f}}^{\rm K}}{\partial\varepsilon}
-\frac{\hbar}{2}
\frac{\partial\underline{f}^{\rm K}}{\partial\varepsilon}
\frac{\partial\underline{\bar{\Delta}}}{\partial t} 
=\underline{0}. 
\label{eq-augmented_Eilenberger_anomalous_velocity-Keldysh-11}
\end{align}
Here, all quantities are written in spin space. Upon taking the trace of Eq.~(\ref{eq-augmented_Eilenberger_anomalous_velocity-Keldysh-11}), integrating over energy, and performing the Fermi-surface average, we find that the spin-orbit commutator vanishes owing to the cyclic invariance of the trace. The electric-field term also vanishes because $\underline g^{\rm K}\to\pm2\underline\sigma_0$ as $\varepsilon\to\pm\infty$ and $\langle{\bf v}_{\rm F}\rangle_{\rm F}={\bf 0}$. The pairing terms cancel upon using the gap equation and its barred counterpart, while the terms containing energy derivatives of $\underline f^{\rm K}$ vanish because $\underline f^{\rm K}\to\underline{0}$ at high energies. Using the gap equation
\begin{align}
\underline{\Delta}
=
\Gamma_0
\int_{-\varepsilon_{\rm c}}^{\varepsilon_{\rm c}}
\frac{d\varepsilon}{4i}
\left\langle
\underline{f}^{\rm K}
\right\rangle_{\rm F}, 
\label{eq-gap-Keldysh}
\end{align}
we obtain
\begin{align}
&i\hbar
\frac{\partial}{\partial t}
\int_{-\varepsilon_{\rm c}}^{\varepsilon_{\rm c}}
d\varepsilon
\left\langle
{\rm Tr}\,\underline{g}^{\rm K} 
\right\rangle_{\rm F} \notag \\
&+i\hbar
{\bm\nabla}\cdot
\int_{-\varepsilon_{\rm c}}^{\varepsilon_{\rm c}}
d\varepsilon
\left\langle
{\rm Tr}
\left\{
\left[
{\bf v}_{\rm F}\underline{\sigma}_0
+
\frac{\partial}{\partial{\bf p}_{\rm F}}
\left(
{\bf g}_{{\bf p}_{\rm F}}\cdot\underline{\bm\sigma}
\right)
\right]
\underline{g}^{\rm K}
\right\}
\right\rangle_{\rm F} \notag \\
&=0. 
\end{align}
The prefactor of the current density is determined by requiring consistency with the expression for the charge density, Eq.~(\ref{eq-rho2}). We then obtain
\begin{subequations}
\begin{align} 
&\frac{\partial\rho}{\partial t}+{\bm\nabla}\cdot{\bf j}=0, \\
{\bf j}
&=
-\frac{eN(0)}{4} \notag \\
&\times
\int_{-\varepsilon_{\rm c}}^{\varepsilon_{\rm c}}
d\varepsilon
\left\langle
{\rm Tr}
\left\{
\left[
{\bf v}_{\rm F}\underline{\sigma}_0
+
\frac{\partial}{\partial{\bf p}_{\rm F}}
\left(
{\bf g}_{{\bf p}_{\rm F}}\cdot\underline{\bm\sigma}
\right)
\right]
\underline{g}^{\rm K}
\right\}
\right\rangle_{\rm F}. 
\label{eq-j-Keldysh} 
\end{align}
\end{subequations}
Equation~(\ref{eq-j}) is obtained from Eq.~(\ref{eq-j-Keldysh}) by analytic continuation to the Matsubara formalism using $\underline{g}^{\rm K}=(\underline{g}^{\rm R}-\underline{g}^{\rm A})\tanh(\varepsilon/2k_{\rm B}T)$ and the residue theorem.



\begin{thebibliography}{9}
\bibitem{ede90} V. M. Edelstein, 
\href{https://doi.org/10.1016/0038-1098(90)90963-C}{Solid State Commun. {\bf 73}, 233 (1990)}. 
\bibitem{esc11} M. Eschrig, 
\href{https://doi.org/10.1063/1.3541944}{Physics Today {\bf 64}, 43 (2011)}.
\bibitem{lin15} J. Linder and J. W. A. Robinson, 
\href{https://doi.org/10.1038/nphys3242}{ Nat. Phys. {\bf 11}, 307 (2015)}. 
\bibitem{ede95} V. M. Edelstein, 
\href{https://doi.org/10.1103/PhysRevLett.75.2004}{Phys. Rev. Lett. {\bf 75}, 2004 (1995)}.
\bibitem{yip02} S. K. Yip, 
\href{https://doi.org/10.1103/PhysRevB.65.144508}{Phys. Rev. B {\bf 65}, 144508 (2002)}.
\bibitem{yip05} S. K. Yip, 
\href{https://doi.org/10.1007/s10909-005-6012-7}{J. Low Temp. Phys. {\bf 140}, 67 (2005)}.
\bibitem{oka06} M. Oka, M. Ichioka, and K. Machida, 
\href{https://doi.org/10.1103/PhysRevB.73.214509}{Phys. Rev. B {\bf 73}, 214509 (2006)}.
\bibitem{kon15} F. Konschelle, I. V. Tokatly, and F. S. Bergeret, 
\href{https://doi.org/10.1103/PhysRevB.92.125443}{Phys. Rev. B {\bf 92}, 125443 (2015)}.
\bibitem{he19} J. J. He, K. Hiroki, K. Hamamoto, and N. Nagaosa,
\href{https://doi.org/10.1038/s42005-019-0230-9}{Commun. Phys. {\bf 2}, 128 (2019)}.
\bibitem{he20} W.-Y. He and K. T. Law, 
\href{https://doi.org/10.1103/PhysRevResearch.2.012073}{Phys. Rev. Research {\bf 2}, 012073 (2020)}. 
\bibitem{dai24}
A. Daido, T. Kitamura, and Y. Yanase,
\href{https://doi.org/10.1103/PhysRevB.110.094505}{Phys. Rev. B {\bf 110}, 094505 (2024)}.
\bibitem{ede03} V. M. Edelstein, 
\href{https://doi.org/10.1103/PhysRevB.67.020505}{Phys. Rev. B {\bf 67}, 020505(R) (2003)}.
\bibitem{fuj05} S. Fujimoto,
\href{https://doi.org/10.1103/PhysRevB.72.024515}{Phys. Rev. B {\bf 72}, 024515 (2005)}.
\bibitem{hay06} N. Hayashi, Y. Kato, P. A. Frigeri, K. Wakabayashi, and M. Sigrist, 
\href{https://doi.org/10.1016/j.physc.2005.12.048}{Physica C {\bf 437-438}, 96 (2006)}. 
\bibitem{ike20} Y. Ikeda and Y. Yanase, 
\href{https://doi.org/10.1103/PhysRevB.102.214510}{Phys. Rev. B {\bf 102}, 214510 (2020)}.

\bibitem{eil68} G. Eilenberger, 
\href{https://doi.org/10.1007/BF01379803}{Z. Phys. {\bf 214}, 195 (1968)}.
\bibitem{yip92} S. K. Yip and J. A. Sauls, 
\href{https://doi.org/10.1103/PhysRevLett.69.2264}{Phys. Rev. Lett. {\bf 69}, 2264 (1992)}. 
\bibitem{xu95} D. Xu, S.-K. Yip, and J. A. Sauls, 
\href{https://doi.org/10.1103/PhysRevB.51.16233}{Phys. Rev. B {\bf 51}, 16233 (1995)}. 
\bibitem{sau22} J. A. Sauls, 
\href{https://doi.org/10.1093/ptep/ptac034}{Prog. Theor. Exp. Phys. {\bf 2022}, 033I03 (2022)}. 
\bibitem{buc95} L. J. Buchholtz, M. Palumbo, D. Rainer, and J. A. Sauls, 
\href{https://doi.org/10.1007/BF00754525}{J. Low Temp. Phys. {\bf 101}, 1079 (1995)}. 
\bibitem{buc95-2} L. J. Buchholtz, M. Palumbo, D. Rainer, and J. A. Sauls, 
\href{https://doi.org/10.1007/BF00754526}{J. Low Temp. Phys. {\bf 101}, 1099 (1995)}.
\bibitem{fog97} M. Fogelstr\"{o}m, D. Rainer, and J. A. Sauls, 
\href{https://doi.org/10.1103/PhysRevLett.79.281}{Phys. Rev. Lett. {\bf 79}, 281 (1997)}.
\bibitem{kle87} U. Klein, 
\href{https://doi.org/10.1007/BF00681621}{J. Low Temp. Phys. {\bf 69}, 1 (1987)}.
\bibitem{sch95} N. Schopohl and K. Maki, 
\href{https://doi.org/10.1103/PhysRevB.52.490}{Phys. Rev. B {\bf 52}, 490 (1995)}.
\bibitem{ich96} M. Ichioka, N. Hayashi, N. Enomoto, and K. Machida, 
\href{https://doi.org/10.1103/PhysRevB.53.15316}{Phys. Rev. B {\bf 53}, 15316 (1996)}.
\bibitem{ich97} M. Ichioka, N. Hayashi, and K. Machida, 
\href{https://doi.org/10.1103/PhysRevB.55.6565}{Phys. Rev. B {\bf 55}, 6565 (1997)}.
\bibitem{vor08} A. B. Vorontsov, I. Vekhter, and M. Eschrig, 
\href{https://doi.org/10.1103/PhysRevLett.101.127003}{Phys. Rev. Lett. {\bf 101}, 127003 (2008)}. 
\bibitem{esc12} M. Eschrig, C. Iniotakis, and Y. Tanaka, in {\it Non-Centrosymmetric Superconductors},
edited by E. Bauer and M. Sigrist (Springer, Berlin, Heidelberg, 2012), p.~313.

\bibitem{kit09} T. Kita, 
\href{https://doi.org/10.1103/PhysRevB.79.024521}{Phys. Rev. B {\bf 79}, 024521 (2009)}. 
\bibitem{uek18} H. Ueki, M. Ohuchi, and T. Kita, 
\href{https://doi.org/10.7566/JPSJ.87.044704}{J. Phys. Soc. Jpn. {\bf 87}, 044704 (2018)}.
\bibitem{mas19} Y. Masaki, 
\href{https://doi.org/10.1103/PhysRevB.99.054512}{Phys. Rev. B {\bf 99}, 054512 (2019)}.
\bibitem{jos20} E. S. Joshua, H. Ueki, W. Kohno, and T. Kita, 
\href{https://doi.org/10.7566/JPSJ.89.104702}{J. Phys. Soc. Jpn. {\bf 89}, 104702 (2020)}.
\bibitem{ohu22} M. Ohuchi, H. Ueki, and T. Kita, 
\href{https://doi.org/10.1103/PhysRevB.105.064514}{Phys. Rev. B {\bf 105}, 064514 (2022)}.
\bibitem{gor59} L. P. Gor'kov, 
Zh. Eksp. Teor. Fiz. {\bf 36}, 1918 (1959) [Sov. Phys. JETP {\bf 9}, 1364 (1959)]. 
\bibitem{gor60} L. P. Gor'kov, 
Zh. Eksp. Teor. Fiz. {\bf 37}, 1407 (1959) [Sov. Phys. JETP {\bf 10}, 998 (1960)].
\bibitem{she18} A. M. Sheikhabadi, I. Miatka, E. Y. Sherman, and R. Raimondi, 
\href{https://doi.org/10.1103/PhysRevB.97.235412}{Phys. Rev. B {\bf 97}, 235412 (2018)}.

\bibitem{nag16} Y. Nagai and H. Nakamura, 
\href{http://doi.org/10.7566/JPSJ.85.074707}{J. Phys. Soc. Jpn. {\bf 85}, 074707 (2016)}.
\bibitem{nag08} Y. Nagai, N. Hayashi, N. Nakai, H. Nakamura, M. Okumura, and M. Machida, 
\href{https://doi.org/10.1088/1367-2630/10/10/103026}{New J. Phys. {\bf 10}, 103026 (2008)}.
\bibitem{uek19} H. Ueki, R. Tamura, and J. Goryo, 
\href{https://doi.org/10.1103/PhysRevB.99.144510}{Phys. Rev. B {\bf 99}, 144510 (2019)}.

\bibitem{ber84} M. V. Berry, 
\href{https://doi.org/10.1098/rspa.1984.0023}{Proc. R. Soc. Lond. A {\bf 392}, 45 (1984)}.
\bibitem{wil84} F. Wilczek and A. Zee, 
\href{https://doi.org/10.1103/PhysRevLett.52.2111}{Phys. Rev. Lett. {\bf 52}, 2111 (1984)}.
\bibitem{xia10} D. Xiao, M.-C. Chang, and Q. Niu, 
\href{https://doi.org/10.1103/RevModPhys.82.1959}{Rev. Mod. Phys. {\bf 82}, 1959 (2010)}.

\bibitem{kit15} T. Kita, {\it Statistical Mechanics of Superconductivity} (Springer, Tokyo, 2015).
\bibitem{wig32} E. P. Wigner, 
\href{https://doi.org/10.1103/PhysRev.40.749}{Phys. Rev. {\bf 40}, 749 (1932)}. 
\bibitem{lon61} F. London, 
{\it Macroscopic Theory of Superconductivity} (Dover, New York, 1961).
\bibitem{yos58} K. Yosida, 
\href{https://doi.org/10.1103/PhysRev.110.769}{Phys. Rev. {\bf 110}, 769 (1958)}. 
\bibitem{kub22} T. Kubo, 
\href{https://doi.org/10.1103/PhysRevApplied.17.014018}{Phys. Rev. Applied {\bf 17}, 014018 (2022)}. 
\bibitem{she20} A. Sheikhzada and A. Gurevich, 
\href{https://doi.org/10.1103/PhysRevB.102.104507}{Phys. Rev. B {\bf 102}, 104507 (2020)}.
\bibitem{hig23} Y. Higashi, S. Yoshizawa, T. Yanagisawa, I. Hase, Y. Mawatari, and T. Uchihashi, 
\href{https://doi.org/10.1103/PhysRevB.108.064504}{Phys. Rev. B {\bf 108}, 064504 (2023)}. 
\bibitem{bok68} J. Bok and J. Klein, 
\href{https://doi.org/10.1103/PhysRevLett.20.660}{Phys. Rev. Lett. {\bf 20}, 660 (1968)}. 
\bibitem{mor71} T. D. Morris and J. B. Brown, 
\href{https://doi.org/10.1016/0031-8914(71)90330-2}{Physica (Amsterdam) {\bf 55}, 760 (1971)}. 
\bibitem{kit10} T. Kita, 
\href{https://doi.org/10.1143/PTP.123.581}{Prog. Theor. Phys. {\bf 123}, 581 (2010)}.
\bibitem{kit01} T. Kita, 
\href{https://doi.org/10.1103/PhysRevB.64.054503}{Phys. Rev. B {\bf 64}, 054503 (2001)}. 
\bibitem{esc99} M. Eschrig, J. A. Sauls, and D. Rainer, 
\href{https://doi.org/10.1103/PhysRevB.60.10447}{Phys. Rev. B {\bf 60}, 10447 (1999)}.
\end{thebibliography}
\end{document}